\documentclass[aps,prc,preprint,showpacs,showkeys,superscriptaddress]{revtex4-1}
\usepackage[T1]{fontenc}
\usepackage{tgtermes}

\usepackage{amsfonts}
\usepackage{amsmath}
\usepackage{amssymb}
\usepackage{graphicx}

\begin{document}

\title{Statistical Properties of $^{56-57}$Fe with Multi-Shell Degrees of
Freedom}
\author{Bassam Shehadeh}
\affiliation{Physics Department, College of Science, Qassim University, Qassim, Burydah
51411, KSA. }
\affiliation{Department of Physics and Astronomy, Iowa State University, Ames, IA 50011,
U.S.A.}
\author{James P. Vary}
\affiliation{Department of Physics and Astronomy, Iowa State University, Ames, IA 50011,
U.S.A.}
\keywords{Nuclear physics, Statistical spectroscopy, Thermal properties of
nuclei, level densities}
\pacs{21.10.Ma, 21.10.Pc, 21.60.Cs, 21.60.n, 24.10.Cn, 24.10.Pa}

\begin{abstract}
We investigate the statistical properties of $^{56-57}$Fe within a model
capable of treating all the nucleons as active in an infinite model space including
pairing effects. Working within the canonical ensemble, our model is built on
single-particle states, thus it includes shell and subshell closures. We include spin correlation parameters to account for
nucleon spin multiplicities. Pairing and shell effects are important in the low
temperature region where the quantum effects dominate. The validity of the
model is extended to higher temperatures, approaching the classical regime, by including continuum states. We present results for excitation
energy, heat capacity and level density of $^{56-57}$Fe. The resulting
favorable comparisons with experimental data and with other approaches, where applicable, 
indicate that the methods we develop are useful for statistical properties
of medium to heavy nuclei.
\end{abstract}

\volumeyear{year}
\volumenumber{number}
\issuenumber{number}
\eid{identifier}
\date[Date text]{date}
\received[Received text]{date}
\revised[Revised text]{date}
\accepted[Accepted text]{date}
\published[Published text]{date}
\startpage{101}
\endpage{102}
\maketitle


\section{Introduction}

Iron nuclei, especially $^{56}$Fe, are the foci of extensive studies within various approaches\cite{ABLNphysrevc15, RHJphysrevc98,NAphyslett98,LANaip00,NAphysrevlett97}
since these nuclei play a major role in astrophysics \cite{NAphysrevlett97}.
The iron nuclei are the heaviest nuclei created by fusion of
charged particles inside stars, and the starting point of the
synthesis of heavier nuclei. Indeed, the statistical properties
of $^{56}$Fe are critical for our understanding of the
fundamental nucleosynthesis processes in iron rich stars and cataclysmic astrophysical events such as neutron star mergers.

A major thrust of the present work is to develop a statistical 
method suitable for reaching higher temperatures by allowing the 
participation of all the nucleons. Another motivation for 
investigating $^{56}$Fe is to compare our methods with another 
modern approach also available for this system. Furthermore,
a comprehensive experimental list of states exists for $%
^{56}$Fe \cite{toi98} as well as experimental level densities for 
$^{56,57}$Fe \cite{SETALphysRev03,SETALNulcInstMethA00}.

Theoretical results for how the excitation energy depends on temperature and how the
level density changes with excitation energy for $^{56}$Fe are available
from Nakada and Alhassid \cite{NAphysrevlett97}. Their results are obtained
using the auxiliary field Monte Carlo shell model or \textquotedblleft Shell
Model Monte Carlo" (SMMC) method. Another version of the Monte Carlo shell
model has also been used by Rombouts, Heyde, and Jachowicz (RHJ) \cite%
{RHJphysrevc98} to obtain $^{56}$Fe statistical properties. Finally, a
simple but useful phenomenological formula, the well-known Bethe formula,
can be used to compare with various results for the level densities. The
parameters of the back-shifted Bethe formula (BBF) for $^{56}$Fe have been
determined \cite{SNphyslettb90,PPphysrevc97}.

In this investigation, we utilize the single-particle (SP) spectra and,
together with a phenomenological treatment of particle-particle correlations,
to develop the properties of the multi-particle system using the canonical
ensemble. The SP spectra can be simple, such as the three-dimensional
harmonic oscillator, or more realistic, such as states obtained from a
phenomenological SP potential \cite{RHJphysrevc98} or provided by
Hartree-Fock (HF) \cite{Pphysrevlett00}. The SP states are then thermally
populated with an approximate treatment of interaction effects to produce
the full system's partition function. Observables, such as excitation
energy, heat capacity, and level densities, can be directly computed. We
call this method the \textquotedblleft Statistical Mechanics using Single
Particle States\textquotedblright (SMSPS).

Our approach models the situation of a specific set of nucleons, for
example those that comprise the $^{56}$Fe nucleus, embedded in a thermal
bath such as the environment of a neutron star or a plasma produced during neutron star mergers. One of our goals is to simulate the physics of excitations that extend beyond those of the valence degrees of freedom alone.  That is, we incorporate core excitations
and excitations into the resonance and continuum regions.  Given this span
of thermal excitations, our approach must remain elementary in several aspects
to make the calculations feasible.

In order to address the nuclear response over a very wide range of temperatures,
we adopt a nuclear Hamiltonian that includes a mean field for all nucleons and
a residual pairing interaction for nucleons occupying valence states.  For
states below the valence states we retain only the mean field Hamiltonian.
Furthermore, we identify an approximate boundary in excitation energy between
resonance states of the mean field and non-resonant continuum states.  The resonant
states are adopted from the quasi-bound states solutions of the mean field.
The non-resonant states are taken to be the single-particle states of a spherical
cavity that are smoothly joined by an average energy shift to the resonant state
spectrum. The continuum state spectrum naturally depends on the volume
(or radius of the spherical cavity) assigned to the nucleus in the thermal medium.  In the case of the neutron star, this volume is controlled by external parameters such as
gravity and rotation rate of the star.

The SMSPS model has two significant advantages: The first advantage
arises, due to the neglect of inter-nucleon interactions, from the exact statistical population of these states which dramatically reduces the computation time. The second advantage is the ability to use a realistic SP spectrum as an input. For example, the SP states will reflect the long-range Coulomb repulsion in the proton degrees of freedom by incorporating the repulsive shift in the proton states relative to the neutron states. This distinct SP spectra, one for protons and one for neutrons, is a dominant
aspect of charge dependence in medium and heavy nuclei.

In this investigation, the SP states provided by Ref.\cite{RHJphysrevc98} are
employed within the SMSPS. In section (\ref{secSM}) we introduce our
approach to the canonical ensemble and present its results. In Section (\ref{secresults}) we present an inter-comparison of our results with experimental data
and with the Bethe formula. Section (\ref{secconclusion}) presents our conclusions and outlook to the future.

\section{Statistical Spectroscopy of The Nuclear System\label{secSM}}

\subsection{The Single Particle Partition Function}

The first basic ingredient of our model is the Hamiltonian which can be written as
\begin{equation}
\hat{H}=\hat{H}_{m}+\hat{H}_{P},  \label{eqham1}
\end{equation}%
where $H_{m}$ is the mean field Hamiltonian, and $H_{P}$ is a Hamiltonian
that accounts for the residual or "effective" interactions that induce
correlations beyond those accommodated by the mean field. For open-shell
nuclear systems, such as $^{56}$Fe, the nuclear pairing Hamiltonian \cite{chasman03} is particularly important. For our specific application, we therefore include pairing interaction contributions to states above the Fermi level.
The second ingredient of our approach is the SP
partition function $\mathcal{Z}_{1}$ defined as 
\begin{eqnarray}
\mathcal{Z}_{1}(\beta ) &=&\underset{\mathcal{Z}_{1}^{\text{SPS}}}{%
\underbrace{\sum_{\alpha ,k}\left\langle \psi _{\alpha }^{(k)}\left\vert
e^{-\beta \hat{H}}\right\vert \psi _{\alpha }^{(k)}\right\rangle }}+  \notag \\
&&\underset{\mathcal{Z}_{1}^{\text{cont}}}{\underbrace{\frac{4\pi V}{h^{3}}
\int_{\sqrt{2mE_{\max }}}^{\infty }e^{-\beta \frac{p^{2}}{2m}}p^{2}dp}}.
\label{eqsppartition0}
\end{eqnarray}
Here, the index $\alpha $ denotes the set of quantum numbers of each, possibly degenerate, state and $k$ labels the degenerate substates. $E_{\max }$ denotes an energy cutoff defining the maximum resonance state of the mean field Hamiltonian, $V$ denotes the volume of continuum state. The first term in the RHS of Eq.(\ref{eqsppartition0})
represents the sum of the discrete single-particle states (SPS) generated by
the mean field Hamiltonian plus the effect of the residual interaction.
The 2$^{\mathrm{nd}}$ term in the RHS of Eq.(\ref{eqsppartition0})
represents the sum of continuum states and the integral can be computed to
yield
\begin{eqnarray}
\mathcal{Z}_{1}^{\text{cont}}(\beta ,V) &=&V\left( \frac{2\pi m}{h^{2}\beta }%
\right) ^{3/2}\times   \notag \\
&&\left( 
\begin{array}{c}
1+2\sqrt{\frac{\beta E_{\max }}{\pi }}e^{-\beta E_{\max }}- \\ 
\mathrm{erf}\left( \sqrt{\beta E_{\max }}\right) 
\end{array}%
\right) .  \label{eqcont1}
\end{eqnarray}
Assuming the adopted mean field has spherical symmetry (which we adopt here) there is degeneracy with respect to angular momentum projection, $m_{J_\alpha}$, the sum over discrete states can be written as
\begin{eqnarray}
\mathcal{Z}_{1}^{\text{SPS}}(\beta )&=&\sum_{\alpha }g_{\alpha }
e^{-\beta\left(\varepsilon_{\alpha }+\langle\hat{H}_P\rangle\right)},\notag \\
&=&\sum_{\alpha }\sum_{m_{J_{\alpha }}=-J_{\alpha}}^{+J_{\alpha }}
e^{-\beta\left(\varepsilon_{\alpha }+\langle\hat{H}_P\rangle\right)},
\label{eqz1sps}
\end{eqnarray}
where $g_{\alpha }=\sum_{k}\left\langle \psi _{\alpha }^{(k)}|\psi _{\alpha
}^{(k)}\right\rangle$ is the degeneracy for the $\alpha ^{\text{th}}$
state. The partition function in Eq.(\ref{eqz1sps}) represents the
partition function of SPS plus pairing. One common description that
links pairing correlations to thermal effects is implemented within the
Bardeen-Cooper-Schrieffer theory. In BCS the pairing Hamiltonian is expressed by
\[
\hat{H}_P=G\sum_\alpha a^{\dagger}_\alpha a^{\dagger}_{\bar{\alpha}}a_\alpha a_{\bar{\alpha}},
\]
where $\alpha$ and $\bar{\alpha}$ denote a state and its time-reversed state,
respectively. To have this model
effective for hot nuclei, a thermal population of particles among the states
is used, mainly by implementing a Monte-Carlo technique
\cite{ABLNphysrevc15,RHJphysrevc98,NAphyslett98,PhysRevC.93.044320}. This technique, however, requires significant computational resources, even for limited orbits in the proximity of the valence shell. Another model for including pairing effects is based on Hartree-Fock-Bogoliubov (HFB) \cite{MER_Physrev03,KVS_NuclPhysA07} theory.
However, the HFB model is limited by particle number fluctuation and quasi-particle parity mixing issues \cite{PhysRevC.88.034324,GLS_PSH_2014}. In our approach, as commonly done in BCS calculations at zero temperature, the pairing interaction is considered acting only on energy levels above the Fermi level. Phenomenologically we can express this effect on the energy levels as
\begin{equation}
\langle\hat{H}_P\rangle =\theta(\varepsilon_f)\Delta(\beta),
\label{epairing}
\end{equation}
where $\theta(\varepsilon_f)$ is the step function
\[
    \theta(\varepsilon_f)= 
\begin{cases}
    1;& \text{if } \varepsilon_\alpha>\varepsilon_f,\\
    0;& \text{if } \varepsilon_\alpha\le \varepsilon_f.
\end{cases}
\]
Thus the SP partition function in Eq.(\ref{eqz1sps}) becomes
\begin{equation}
\mathcal{Z}_1^{\text{SPS}}(\beta)=\sum_{\alpha}g_\alpha \,
e^{-\left[\varepsilon_\alpha+\theta(\varepsilon_f)\Delta(\beta)\right]\beta}.
\label{eqZ1P2}
\end{equation}

The effect of the pairing is to increase the energy gap of the Fermi level.
The gap shift $\Delta$ is set to be temperature dependent, expected to be maximum near
zero temperature and to decrease as temperature increases to reach zero at a higher
temperature. The calculation of $\Delta$ will be discussed in sec.\ref{secresults}.

Within the framework of the SMSPS the effects of pairing and spin-orbit splitting are averaged by arranging the nucleons using a technique, which we call configuration-restricted recursion. We implement this technique for a fixed species of protons
or neutrons by constructing the partition function for a subgroup of that
species that, for example, limits the values of the magnetic projection of the total
angular momentum quantum number to positive values, signified by $|m_J|$. Thus, we obtain the SPS partition function of that restricted range, denoted as
$\left( \mathcal{Z}_{1}^{\text{SPS}}\right) ^{\left\vert m_{J}\right\vert }$, and given by
\begin{equation}
\left[\mathcal{Z}_{1}^{\text{SPS}}(\beta)\right]^{\left\vert m_{J}\right\vert
}=\sum_{\alpha }\sum_{m_{J_{\alpha }}=J_{\alpha },J_{\alpha }-1,\ldots
>0}\exp \left( -\beta E_{\alpha }\right),  \label{eqz1spsres}
\end{equation}
where $E_\alpha=\varepsilon_\alpha+\theta(\varepsilon_f)\Delta(\beta)$.
With symbol \textquotedblleft $-\left\vert m_{J}\right\vert $\textquotedblright , we signify that magnetic state summation in the partition function
$\left( \mathcal{Z}_{1}^{\text{SPS}}\right) ^{-\left\vert m_{J}\right\vert }$
runs over negative values $\left( -J_{\alpha },-J_{\alpha
}+1,\ldots <0\right) $. Making use of Eq.(\ref{eqcont1}) and Eq.(\ref{eqz1spsres}) 
within Eq.(\ref{eqsppartition0}) we obtain $\mathcal{Z}_{1}^{\left\vert m_{J}\right\vert}$
\begin{equation}
\mathcal{Z}_{1}^{\left\vert m_{J}\right\vert }(\beta)=\left[\mathcal{Z}_{1}^{\text{%
SPS}}(\beta)\right]^{\left\vert m_{J}\right\vert }+\mathcal{Z}_{1}^{\text{cont}%
}(\beta ,V).  \label{eqz1res}
\end{equation}

\subsection{The Nuclear Partition function and Observable Calculations}

We then recur the partition function in Eq.(\ref{eqz1res}) for the given $m$-projected
configurations of $Z_1$ protons (or $N_1$ neutrons) using the recursion
formula for spinless identical fermions, given in Ref.\cite{BFjcp93}. This formula
for $n$ identical spinless (or polarized) fermions is
\begin{equation}
\mathcal{Z}_{n}(\beta )=\frac{1}{n}\sum_{\nu =1}^{n}(-1)^{\nu +1}\mathcal{Z}%
_{1}(\nu \beta )\mathcal{Z}_{n-\nu }(\beta ),\qquad \mathcal{Z}_{0}(\beta)=1,
 \label{eq9}
\end{equation}
where $n$ has to be replaced by $Z_1$ for the proton case ($N_1$ for the neutron case)
in the $|m_J|$ states and by $Z_2$ for the proton case ($N_2$ for the neutron case)
in the $-|m_J|$ states. We denote $Z=Z_1+Z_2$ as the total number of protons, $N=N_1+N_2$ as the total number of neutrons, and $A=Z+N$ as the total number of nucleons. Note that for even $Z$ (or $N$) then $Z_1=Z_2=Z/2$ (or $N_1=N_2=N/2$). For odd $Z$ (or $N$) we choose $Z_1=(Z+1)/2$ and $Z_2=(Z-1)/2$ (or $N_1=(N+1)/2$ and $N_2=(N-1)/2$).

The recursion formula (\ref{eq9}) gives exact values of the
partition functions at specific values of $n$ and $\beta$. Up to this point we obtain
$\mathcal{Z}_{Z_1}^{\left\vert m_{J}\right\vert}$ and $\mathcal{Z}_{N_1}^{\left\vert
m_{J}\right\vert}$ for $Z_1$ protons and  $N_1$ neutrons, respectively, which thermally
populate the $|m_J|$'s states. Following the corresponding procedure, we now obtain
$\mathcal{Z}_{Z_2}^{-\left\vert m_{J}\right\vert}$ and $\mathcal{Z}_{N_2}^{-\left\vert
m_{J}\right\vert}$ for $Z_2$ protons and $N_2$ neutrons that populate the $-|m_J|$'s
states.

The total nuclear partition function is constructed by
\begin{equation}
\mathcal{Z}_{A}(\beta )=\left(\mathcal{Z}_{Z_1}^{\left\vert
m_{J}\right\vert }\mathcal{Z}_{Z_2}^{-\left\vert m_{J}\right\vert
}\right) \left(\mathcal{Z}_{N_1}^{\left\vert
m_{J}\right\vert }\mathcal{Z}_{N_2}^{-\left\vert m_{J}\right\vert
}\right),
\label{CRReq1}
\end{equation}
With this product in Eq.(\ref{CRReq1}) of restricted sub-configuration partition functions,  the total magnetic projection quantum number is restricted, on average, to zero for each nucleon species (protons or neutrons). Thus, the fact that each nucleon species occupies SP states predominantly in spin-paired configurations is taken into account, on average.

Eq.(\ref{CRReq1}) also implies that, at sufficiently high $\beta $, due to the
degeneracy of time-reversed single particle states, for each proton (or
neutron) in state $\left\vert J_{\alpha }^{(Z_1)},|m_{J_\alpha}^{Z_1}|\right\rangle$
(or $\left\vert J_{\alpha }^{(N_1)},|m_{J_\alpha}^{N_1}|\right\rangle$), there exists
another proton (or neutron) in state $\left\vert J_{\alpha
}^{(Z_2)},-|m_{J_\alpha}^{Z_2}|\right\rangle $ (or $\left\vert J_{\alpha
}^{(N_2)},-|m_{J_\alpha}^{N_2}|\right\rangle $) with the same statistical weight. Then,
as $\beta $ decreases, excitations enter that break this symmetry while
retaining only a total $M=0$ constraint for each nucleon species. We believe that this feature of our total partition function is reasonable since we expect that polarization effects are minimal at zero temperature and those polarization effects beyond simple thermal excursions that are accommodated here are unlikely to contribute as temperature increases.

In this way, we approximate average pairing of total spin projections suitable for the
low to moderate excitation region. The additional coherent pairing energy associated
with the ground state and lowest-lying states is very important when we calculate
the level density for an open-shell even-even nucleus. This will be addressed
phenomenologically when calculating level densities via the parameter $\Delta$ in
Eq.(\ref{epairing}).
The configuration-restricted recursion technique and the coherent pairing
energy shift defines our proton-proton and neutron-neutron correlation
approach in our SMSPS theory.

Once the nuclear partition function is computed for the desired system at a given
temperature, the observables such as the average thermal energy $E_{A}$,
the heat capacity $C_{A}$, and the number of levels per unit energy $g_{A}$
can be computed, respectively, in the canonical ensemble as \cite{FEW_AstroJour78}
\begin{equation}
E_{A}(\beta )=-\frac{\partial }{\partial \beta }\log Z_{A}(\beta );\quad
C_{A}(\beta )=-k_{B}\beta ^{2}\frac{\partial E_{A}}{\partial \beta },
\label{eq18}
\end{equation}
and
\begin{eqnarray}
g_{A}(E)&=&\frac{\beta e^{S(\beta )/k_{B}}}{\sqrt{2\pi C_{A}(\beta )/k_{B}}},\notag \\
&=&\frac{\beta }{\sqrt{2\pi C_{A}(\beta )/k_{B}}}Z_{A}(\beta )e^{\beta
E_{A}(\beta )}.  \label{eq33}
\end{eqnarray}
The factor $Z_{A}(\beta )\exp \left( \beta E_{A}(\beta )\right) $ is the
number of microstates $\Omega _{A}(E)$. We also notice that the heat
capacity and the level densities are independent of the lowest-state energy $%
E_{\mathrm{gs}}$ as expected. We can easily prove in the canonical ensemble that \cite%
{phd}%
\begin{equation}
\Delta E\equiv \sqrt{\left\langle E_{A}^{2}\right\rangle -\left\langle
E_{A}\right\rangle ^{2}}=\frac{1}{\beta }\sqrt{C_{A}(\beta )/k_{B}}.
\label{statvarianceEq}
\end{equation}%
Therefore Eq.(\ref{eq33}) can be written as%
\begin{equation*}
g_{A}(E)=\frac{1}{\sqrt{2\pi }}\frac{\Omega (E)}{\Delta E}.
\end{equation*}%
As $\beta \rightarrow \infty $, $\Delta E\rightarrow 0$ and the statistics
become very low. Thus, at $T=0$ we switch over to the micro-canonical ensemble
to obtain level densities \cite{phd}.

\section{Results and Discussion\label{secresults}}

\begin{figure}[ptbh]
\begin{center}
\includegraphics[width=0.90\textwidth]{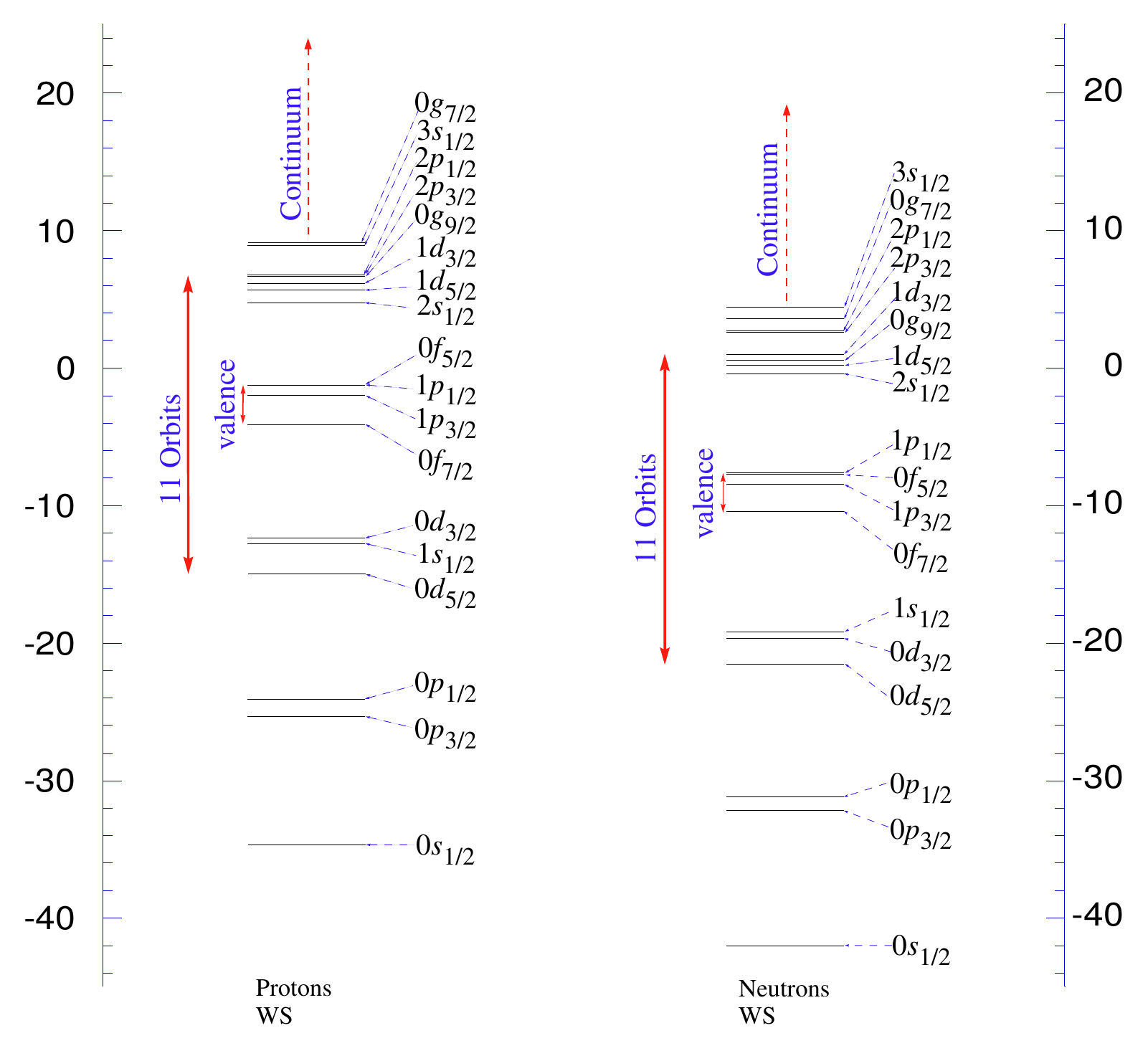}
\end{center}
\par
\caption{(Color online) The single-particle (SP) energy levels in MeV of $^{56,57}$Fe from Wood-Saxon (WS) calculations of Ref.\cite{RHJphysrevc98} indicated as horizontal bars.  For the proton levels, the Coulomb potential of a uniformly charged sphere is included.  Vertical arrows indicate the level groupings used in the calculations.  The continuum region is added smoothly to the discrete spectra as discussed in the text.}
\label{SPS_rambouts}
\end{figure}

Our investigation falls into two main categories: First the effect of model space
on observable quantities such as excitation energy $E_x$, heat capacity $C$, and
level density $\rho$ of $^{56,57}$Fe. Second, the effect of pairing on SP states
and how incorporating this effect would improve level densities $\rho$ versus
experimental results. In both categories, We use SPS of the mean field
spectrum of the Woods-Saxon (WS) potential with phenomenological parameters as
described in Ref.\cite{RHJphysrevc98} to evaluate the thermal properties of $^{56,57}$Fe.
The SPS energy levels and model space divisions are shown in Fig(\ref{SPS_rambouts}).
To be able to reach the lowest possible temperature for larger model spaces we implement
a multi-precision algorithm called \textquotedblleft quad double" developed by Hida
and Bailey \cite{hida1,hida2} which enables us to compute observables up to 212 bits of
floating-point accuracy. At very low temperature and large model space, we implement
ARbitrary PRECision Computation Package (ARPREC) \cite{arprec} which enables us to compute
at 200 digits after the decimal point. We are able to obtain stable results for the canonical ensemble down to $T=1.37$ MeV for our no-core model space.

\subsection{The Model Space Effect}

\begin{figure}[ptbh]
\begin{center}
\includegraphics[width=0.90\textwidth]{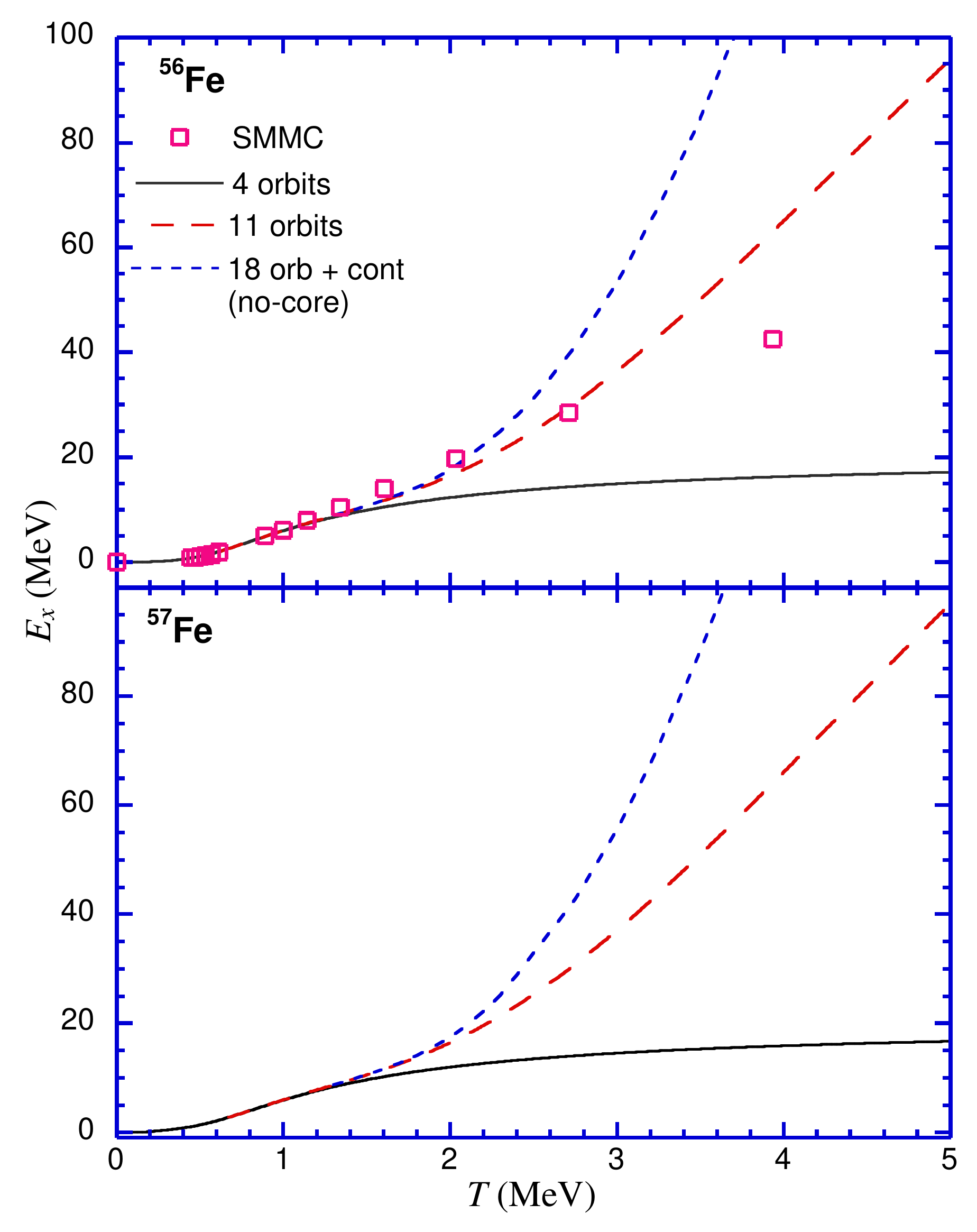}
\end{center}
\par
\vspace{-1.0cm}
\caption{(Color online) The SMSPS excitation energies (in MeV) versus temperature
(in MeV) with pairing set to zero for $^{56}$Fe (top) and  $^{57}$Fe (bottom) computed at various model spaces, 4 
orbits, 11 orbits and 18 orbits+cont (the no-core case) (see common legend). The results for $^{56}$Fe are compared 
with those of the SMMC in Ref.\protect\cite{NAphysrevlett97}.}
\label{E_lowT}
\end{figure}

\begin{figure}[ptbh]
\begin{center}
\includegraphics[width=0.90\textwidth]{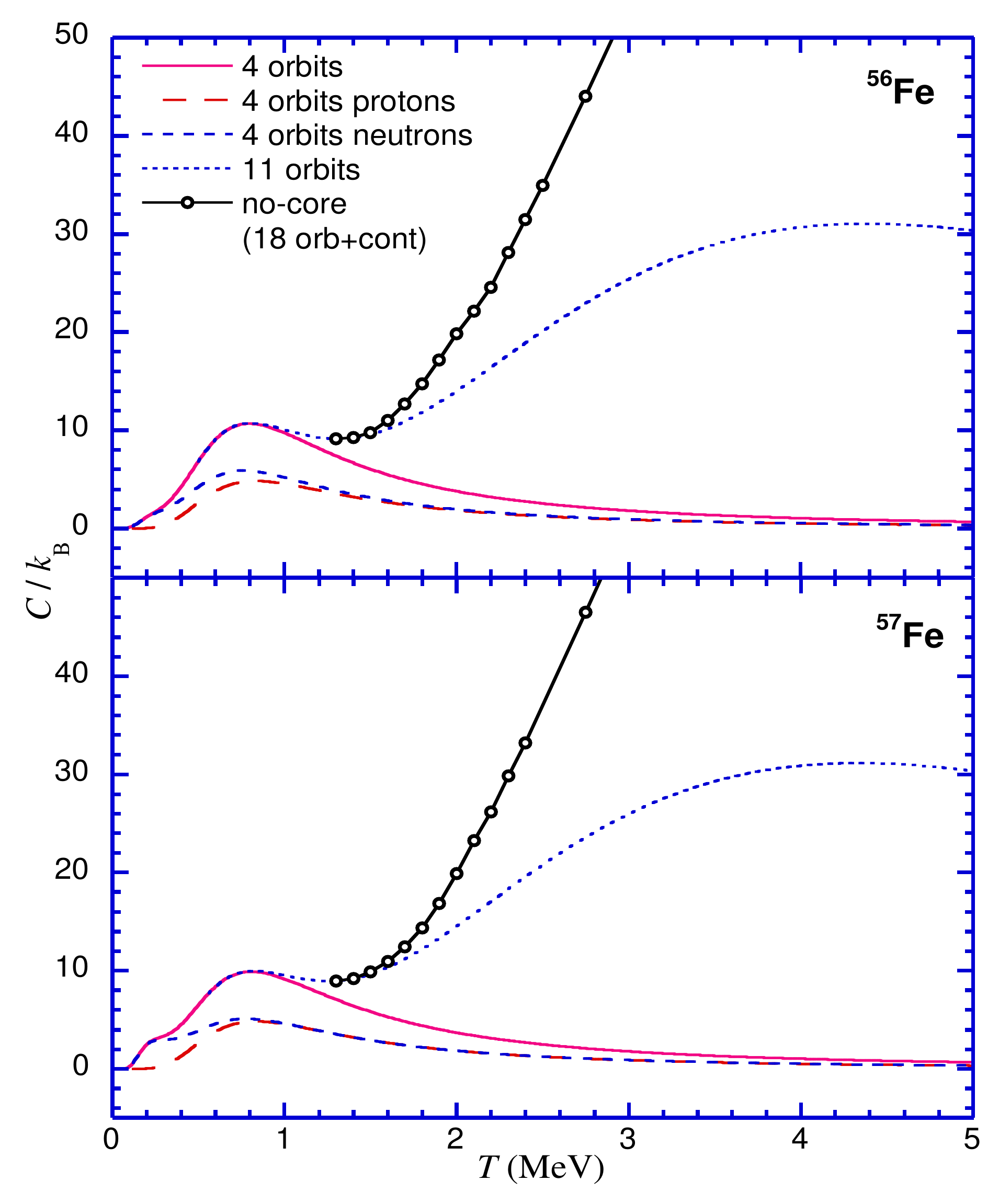}
\end{center}
\par
\caption{(Color online) The SMSPS Heat capacity versus temperature (in
MeV) with pairing set to zero for $^{56,57}$Fe for three model spaces 4 orbits, 11 orbits, and 18 orbits+cont
(no-core). The legend is common for the top and bottom figures.}
\label{C_lowT}
\end{figure}

\begin{figure}[ptbh]
\begin{center}
\includegraphics[width=0.90\textwidth]{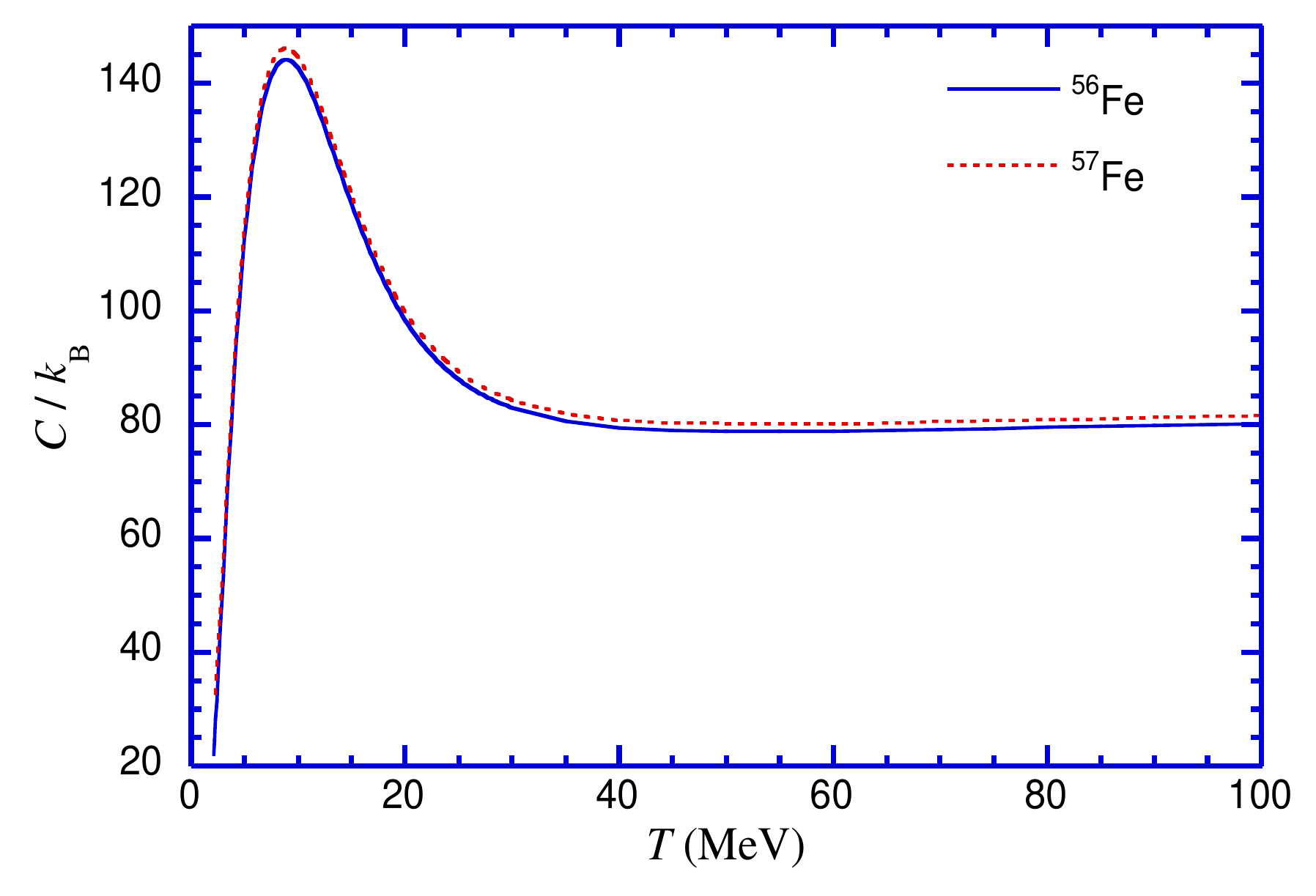}
\end{center}
\par
\caption{(Color online) The SMSPS heat capacities versus temperature (in MeV) with pairing set to zero for $^{56,57}$Fe in 18 orbits+cont (no-core) model space. The continuum is computed for a sphere with radius equal to 2.3 times the root mean square nuclear radius.}
\label{C_HiT}
\end{figure}

\begin{figure}[ptbh]
\begin{center}
\includegraphics[width=0.90\textwidth]{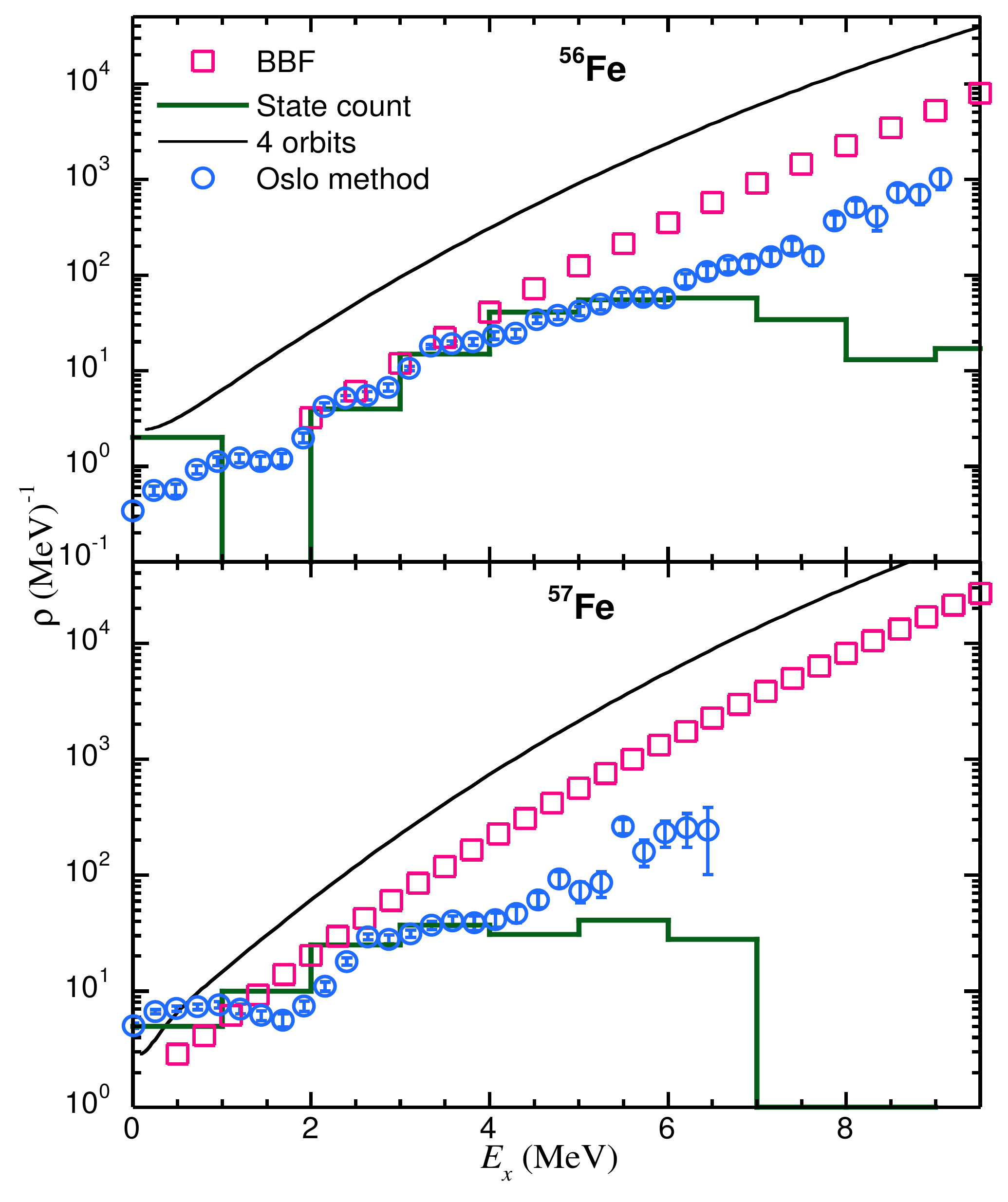}
\end{center}
\par
\caption{(Color online) The SMSPS level density of $^{56,57}$Fe with pairing set to zero computed using 4 orbits (valence shell) is compared with various results. The histogram is the experiment. The BBF level density is evaluated using Eq.(\protect\ref{eqBBF1}) with backshift parameter $\delta=1.38$ MeV \protect\cite{WFHZatomnucldatatab78} and $a=8$ MeV$^{-1}$ for $^{56}$Fe, and $\delta=0$ MeV and $a=8.143$ MeV$^{-1}$ for $^{57}$Fe. The top and bottom figures have common legend.}
\label{g_lowE}
\end{figure}

\begin{figure}[hptb]
\begin{center}
\includegraphics[width=0.90\textwidth]{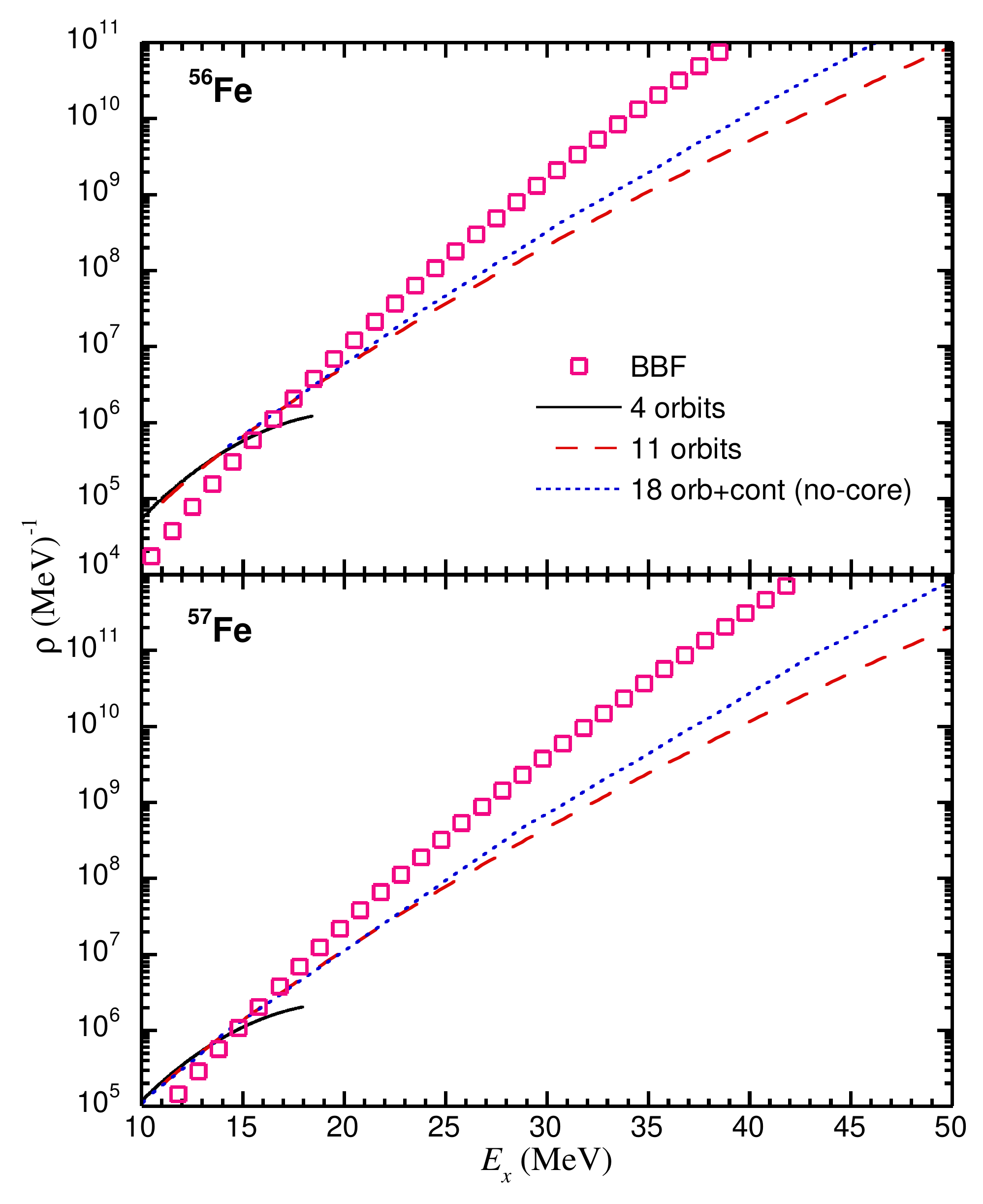}
\end{center}
\par
\caption{(Color online) The level densities with pairing set to zero for $^{56,57}$Fe calculated at higher
excitations at larger model spaces see legend.}
\label{g_EHi}
\end{figure}

In this subsection (covering results presented in Figs.\ref{E_lowT}-\ref{g_EHi}), we set our pairing shift parameter $\Delta=0$. We carry out the
computations for three divisions of the model spaces, shown in
Fig.(\ref{SPS_rambouts}). These three choices allow us to investigate the model space dependence of results over a range of temperatures.
Starting from a simple valence 4 orbits $0f_{7/2}1p_{3/2}1p_{1/2}0f_{5/2}$ which
contains 6 protons and 10 neutrons for $^{56}$Fe (11 neutrons for $^{57}$Fe) leaving
an inactive core of 20 protons and 20 neutrons. We simply refer to this model space as
4 orbits. The second model space consists of 18 protons and 22 neutrons for $^{56}$Fe
(23 neutrons for $^{57}$Fe) in 11 orbits, and runs from the $0d_{5/2}$ state to the$0g_{9/2}$ state, leaving a core of 8 protons and 8 neutrons. We refer to this model space as 11
orbits. The final model space is the no-core case which includes all available orbits
(18 orbits) and continuum to attain a no core system. The volume of the continuum state
is considered to be a spherical cavity with a radius equal to 2.3 times the mean nuclear
radius. We adopt a mean nuclear radius $R=r_0A^{1/3}$ where $r_0=1.25$ fm. Thus $R=4.78$ fm for $^{56}$Fe and $R=4.81$ fm for $^{57}$Fe. Support for this choice of volume for the continuum states will be presented when we discuss the heat capacity.

Fig.(\ref{E_lowT}) shows the excitation energy of $^{56,57}$Fe versus temperature
computed with SMSPS using 4-orbit, 11-orbit, and no-core model spaces. Here,
the excitation energies of the SMSPS are compared with the excitation energy obtained
for $^{56}$Fe using the Shell-Model Monte Carlo (SMMC) method of Nakada and Alhassid
\cite{NAphysrevlett97}. The SMMC is a standard technique since it includes a microscopic
Hamiltonian that accounts for important residual interactions.

The general trend for the excitation energy is that it is zero at and near zero
temperature and it does not exhibit any significant increase for $T<0.2$ MeV.
This indicates the thermal energy is insufficient to excite nucleons to
higher valence states. As temperature increases above $T=0.2$ MeV, internal energy
increases rapidly with temperature. The rate of rising excitation energy with
temperature tapers off for 4 orbits and then for 11 orbits model spaces, as the model
space eventually cuts off and yields no additional higher-energy states for the excited
nucleons to occupy. That is, the internal energy saturates towards the maximum energy
offered by the quantum configurations of these limited model spaces. Thus, above a certain temperature a larger model space is needed to provide more particles and states to contribute to the excitation energy.

The SMSPS 4 orbits results agree well with those of the SMMC up to $T\leq 1$ MeV. The results of the SMSPS for $T\leq 1$ MeV are controlled by phenomenological SP spectra that are in good agreement with experimental information. Beyond the 1 MeV temperature, the SMSPS results lie below those of SMMC even for higher model spaces. Note that the SMMC results are obtained with an additional orbital, the $0g_{9/2}$, included within its valence space. We observe that when we include this orbital and additional orbitals in the 11 orbit space, we extend the temperature range over which our SMSPS results agree with the SMMC results. We also observe that the SMSPS internal energy is close to the internal energy predicted in Ref.\cite{RHJphysrevc98} using the shell-model quantum Monte Carlo method and pairing strength $G=20$ MeV/56 which we attribute to our use of the same SP spectra.

To expand the range of applicability of the SMSPS, especially for $T\geq 1$ MeV, we increase the model space to 18 orbits with the continuum (no-core). For the no-core model space we are able to extract reliable results when $T\geq 1.5$ MeV by implementing high precision algorithms of Refs.\cite{hida1,hida2,arprec} within the SMSPS code. Going lower in temperature with no core model space is possible with increasing number of significant digits of the code's variables. The computation time, however, becomes lengthy. Thus we truncate the minimum temperature limit for larger model spaces such that the results of larger model spaces join smoothly with the results in the smaller model spaces.

In the case of $^{57}$Fe, there are 11 neutrons in the $fp$ shell at T = 0. Therefore, this isotope has a leading unperturbed configuration which is one neutron less
than needed to fill the $1p_{3/2}$ subshell. We found that the addition of
one neutron to $^{56}$Fe does not have a significant contribution to the
excitation energy as a function of $T$ as seen by comparing results in the two panels of Fig.(\ref{E_lowT}). Although it is not apparent from Fig.(\ref{E_lowT}), for $E_{x}\geq1$ MeV in the 4 orbit model space
the $^{56}$Fe excitation energy as a function of $T$ is slightly greater than that of $^{57}$Fe. This apparent paradox arises from the fact that, in this limited model space the ratio of the number of accessible states for excitation to the number of nucleons is sufficiently smaller for $^{57}$Fe. Therefore, there are fewer degrees of freedom for excitation for the neutrons for $E \geq 1$ MeV in $^{57}$Fe and it is closer to saturating the 4 orbit model space than is $^{56}$Fe.

Fig.(\ref{C_lowT}) displays the SMSPS heat capacity for $^{56,57}$Fe as
a function of temperature for the three model spaces, 4 orbits, 11 orbits, and no-core
as described in Fig.(\ref{SPS_rambouts}). For the 4-orbit model space, the maximum heat capacity occurs when one starts exciting nucleons from the inner valence state (the
$0f_{7/2}$ state). The neutron heat capacity at the peak is somewhat larger than the 
proton's reflecting the larger number of neutron degrees of freedom in this model space.

Because there is a small energy spacing between the $1p_{3/2}$ and the
$0f_{5/2}$ single-neutron states (0.78 MeV) the neutron heat capacity is
larger than the proton's in the low temperature region for both $^{56}$Fe
and $^{57}$Fe. In fact, there is a small hump in the total heat capacity
which comes entirely from the neutron heat capacity at $T=0.2$ MeV, barely
visible in Fig.(\ref{C_lowT}) for $^{56}$Fe but more apparent for $^{57}$Fe.
This signifies the excitation from the $1p_{3/2}$ state to the $0f_{5/2}$ and
$1p_{1/2}$ states since they are nearly degenerate. By contrast, the lowest proton
excitation is more than 2 MeV from the $0f_{7/2}$ to the $1p_{3/2}$ state.
The SMSPS heat capacity for $^{57}$Fe, where we have 3 neutrons in the $1p_{3/2}$
instead of 2 neutrons in the $^{56}$Fe case, shows an enhanced hump at $T=0.2$ MeV.

Both $^{56,57}$Fe heat capacities exhibit a prominent peak in heat capacity
near $T=0.8$ MeV. The expanding model spaces, exhibited for $^{56,57}$Fe, show that this feature survives as the core degrees of freedom are brought into the calculation. We therefore identify this major heat capacity peak as an indicator of what we call
\textquotedblleft Total Valence Melting\textquotedblright (TVM) since it signifies thermally exciting all the valence nucleons from their $T=0$ subshells.

The core contribution to heat capacity above the TVM temperature is very
important. Any consideration of the heat capacity beyond TVM temperature
must include the inner-shell nucleons and higher lying orbits. Below the TVM
temperature, however, all model spaces produce exactly the same behavior.
This is understandable since at sufficiently low temperatures the valence
nucleons dominate. We conclude that valence shell contributions to the excitation
energy and heat capacity are dominant for $T<1.0$ MeV.

Fig.(\ref{C_HiT}) shows the SMSPS heat capacity for $^{56,57}$Fe computed at no core model space including the continuum for an extended temperature range. There is another maximum for heat capacities at $T=8.8$ MeV, which marks the transition from resonance states to the continuum. The value and the position of this maximum depend on the continuum state volume: the larger the volume - the larger the maximum. As we vary the volume, the position of the maximum versus temperature can shift either towards higher temperature or towards lower temperature. We fix the volume such that the maximum is located at a temperature corresponding to the binding energy per nucleon. In case of $^{56,57}$Fe $BE/A\approx 8.8$ MeV we find there is a unique value for the spherical volume of the continuum which satisfies this condition - when the radius of the volume equals to 2.3 mean nuclear radius of $^{56,57}$Fe.

The rise of the value of the maximum when the volume increases, implies that the system finds it more favorable, as expected, to excite nucleons  from resonance states to continuum states. We note that this maximum corresponds to the beginning of a phase transition from quantum gas to classical gas regime since the values of the heat capacities $C/k_B$ are trending towards the classical value $\frac32A$ ($84$ for $^{56}$Fe and 85.5 for $^{57}$Fe).

Fig.(\ref{g_lowE}) displays the level densities at low excitation energies of 
$^{56,57}$Fe predicted by the SMSPS and compared with results from other methods and with experiment where available. The SMSPS is computed for valence shells only. As we
can understand from $E_x$ vs $T$ and $C$ vs $T$ graphs the valence contribution is 
sufficient for $T<1.0$ MeV or $E_x<6.0$ MeV. The experimental level density is 
histogrammed into 1 MeV bins using the state list in Ref.\cite{toi98}. The experimental 
spectrum appears incomplete for $E_x>4.0$ MeV. The circular points in the figure
represent the experimental level densities extracted from primary $\gamma$ spectra for
$^{56,57}$Fe nuclei obtained with ($^3$He, $\alpha\gamma $) and ($^3$He, $^3$He$\gamma$)
reactions on $^{56,57}$Fe targets using the Oslo method (OM) \cite{SETALphysRev03,
SETALNulcInstMethA00}.

The Bethe backshifted formula values (BBF) displayed in Fig.(\ref{g_lowE}) are 
computed using \cite{SNphyslettb90,PPphysrevc97} 
\begin{equation}
g(E_{x})=g_{0}\frac{\sqrt{\pi}}{24\sigma}a^{-\frac{1}{4}}\left(
E_{x}-\delta\right) ^{-\frac{5}{4}}\exp\left( 2\sqrt{a\left(
E_{x}-\delta\right) }\right) ,  \label{eqBBF1}
\end{equation}
where $g_{0}=\sqrt{2/\pi}$ and $\sigma$ is the spin cut-off parameter given
by \cite{GCcanjphys65}%
\begin{equation}
\sigma^{2}=0.0888A^{2/3}\sqrt{a\left( E_{x}-\delta\right) }.  \label{eqBBF2}
\end{equation}
For $^{56}$Fe, we choose the backshift parameter $\delta=1.38$ MeV \cite
{WFHZatomnucldatatab78} and the value $a=A/K$ \cite{NHMMQSWWphysrevc02}
where $K=7$ MeV. The same parameter values are employed for $^{57}$Fe, except for the backshift parameter $\delta=0$ MeV. This yields a level density that agrees very well with the experimental state counts and OM level density points for $E_{x}<4.0$ MeV.
At higher excitation energies, we expect the theory to be higher than the experiment
as the reaction mechanism becomes less successful in reaching states of higher angular 
momentum.

Without pairing, the SMSPS results for $^{56}$Fe overestimate level density, almost systematically. If the same backshift parameter used in the BBF ($\delta=1.38$ MeV) is applied to the SMSPS model, the resulting curve would describe well the experimental results at low excitations. This backshift value is an approximate correction to estimate the coherent pairing energy shift which we expect will arise when we include our treatment of pairing in the next subsection.

A similar behavior of SMSPS level density without pairing is obtained for $^{57}$Fe but with a somewhat better agreement with BBF and experiment. The addition of the odd neutron in the valence shell lowers the deficit from the missing pairing contributions which in turns improves the agreement between SMSPS values (absent pairing) with experiment and BBF's. We also notice in Fig.(\ref{g_lowE}) the level density of $^{57}$Fe is higher than that of $^{56}$Fe since the extra neutron is unpaired and can be excited easily to higher states. For $E_x>15.5$ MeV SMSPS level densities at any model space, as shown in Fig.(\ref{g_EHi}), lie lower than the BBF values. Thus, the SMSPS density of states suggest a slower exponential increase with excitation energy than that obtained with the BBF using conventional parameter selections.

To investigate the model space dependence of the level density prior to including pairing effects, we calculate the SMSPS level density for $^{56,57}$Fe at 11 orbits and 18 orbits+continuum (no-core) as shown in Fig.(\ref{g_EHi}). Model space limits inhibit the rate of increase in the level density. In fact, every finite model space produces a curve that
reaches some saturated value as seen with the 4 orbit level density in
Fig.(\ref{g_EHi}). In the figure all level density curves for different
model spaces coincide for $E_{x}<15$ MeV where the 4-orbit model space
(valence orbits only) is adequate to describe the system's level density.
Above 15 MeV it appears safer to adopt a larger model space of at least 11
orbits. For $E_{x}>20$ MeV we need a no-core model space to describe the level
density. Above 21 MeV the effect of the continuum is imperative and has to be
incorporated into the calculation for reliable results.

\subsection{The Pairing Effect}

\begin{figure}[hptb]
\begin{center}
\includegraphics[width=0.90\textwidth]{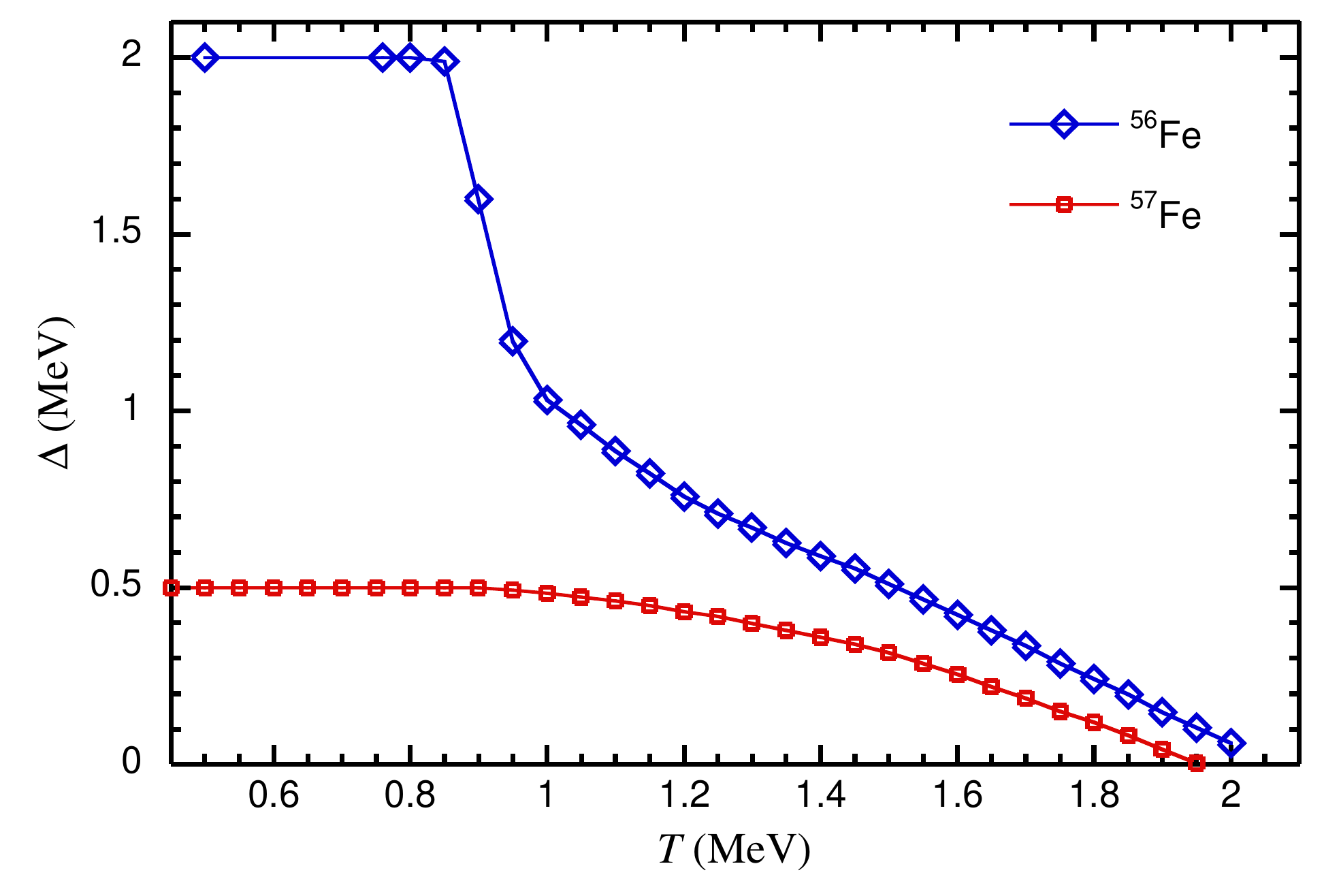}
\end{center}
\par
\caption{(Color online) the pairing energy gap $\Delta$ as a function of temperature
$T$ for $^{56}$Fe and $^{57}$Fe.}
\label{D_T}
\end{figure}

We have commented that shifting the excitation energies by 1.38 MeV (the same energy
shift used in BBF) improves SMSPS level density for $^{56}$Fe towards reproducing the experimental and BBF values. Until now, we have not incorporated our treatment of pairing.  Hence, we now adjust the energy gap of Fermi levels to obtain the shifted
energy $E_x+1.38$ MeV at the same temperature $T$ that originally gives $E_x$. We
perform this energy shift to cover all excitation energies and obtain sufficient data
for $\Delta$ vs $T$. Fig.(\ref{D_T}) shows the resulting pairing gap energy $\Delta$ as a function
of temperature for $^{56}$Fe and $^{57}$Fe.

For $^{56}$Fe, when $T\le 0.885$ MeV the value of $\Delta=2.01$ MeV is constant. Just
when $T>0.885$ MeV, the value of $\Delta$ drops very rapidly until $T=1$ MeV. For $T>1$
MeV, $\Delta$ starts decreasing slowly with increasing $T$. In order to capture the
correct $T$ at which $\Delta$ disappears, we computed the SMSPS for 11 orbits (18 protons
and 22 neutrons) leaving only 8 protons and 8 neutrons as the inert core. The value of
$\Delta$ diminishes at $T=2.065$ MeV. A much smoother behavior is noticed for $^{57}$Fe.
when $T\le 0.906$ MeV, the value of $\Delta=0.501$ MeV is constant. When $T>0.906$ MeV
$\Delta$ decreases smoothly as $T$ increases. This is an indication that the thermally active nucleons of $^{56,57}$Fe are in a superfluid phase below these critical temperatures.
This is similar to what Ref.\cite{PhysRevC.88.034324,GLS_PSH_2014} has concluded for $^{161,162}$Dy and $^{171,172}$Yb isotopes. The generated values of $\Delta$ are fitted as a function of $1/T$ using a 5th-degree polynomial and the fitting coefficients are fed into the SMSPS code to re-evaluate the partition functions and all of the observables at given temperature and energy gap $\Delta$.
 
\begin{figure}[hptb]
\begin{center}
\includegraphics[width=0.90\textwidth]{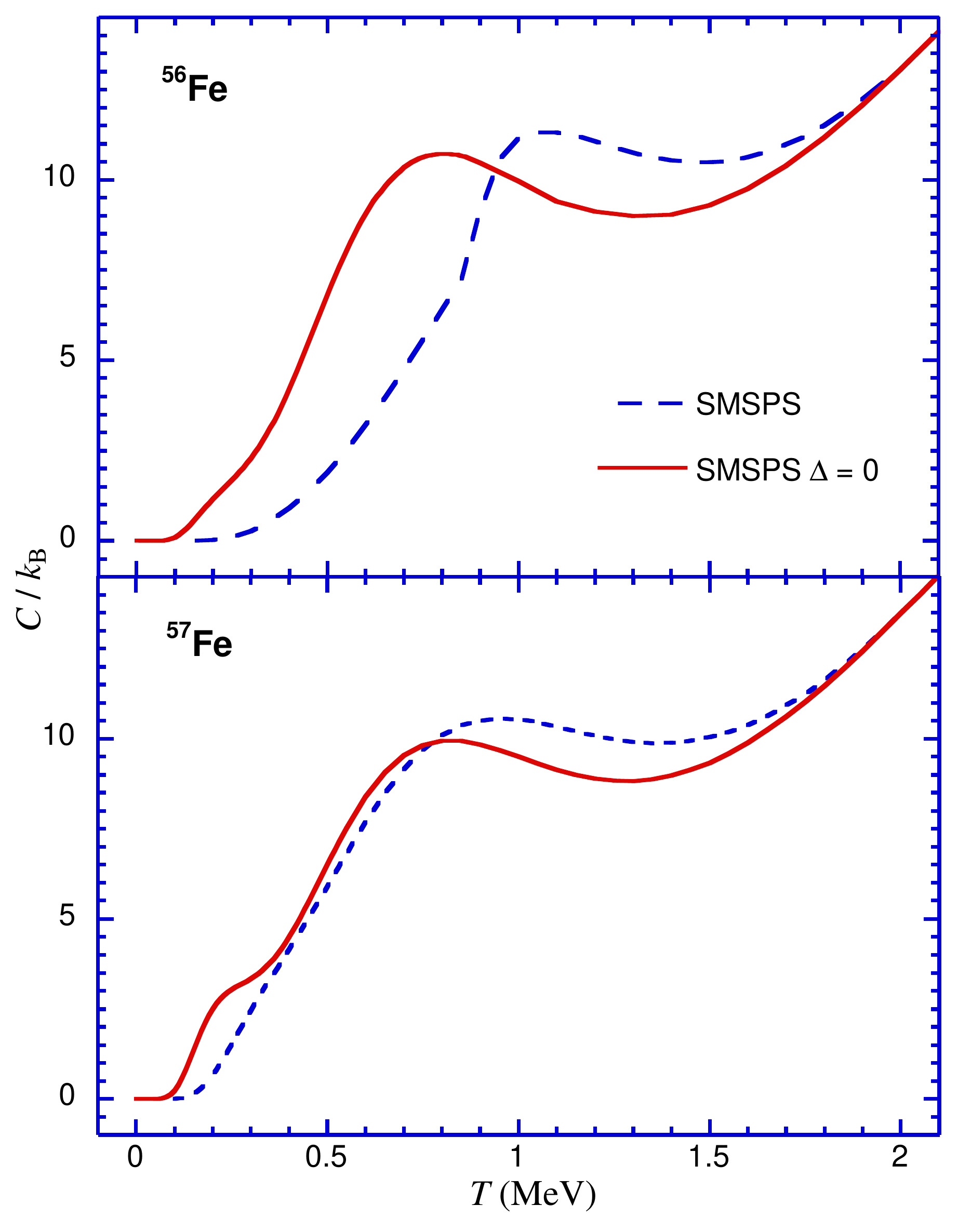}
\end{center}
\par
\caption{(Color online) The effect of the pairing on SMSPS heat capacities versus temperature (in MeV) for $^{56}$Fe (top) and $^{57}$Fe (bottom). All SMSPS results are computed at 11-orbit model space.}
\label{C_Tpairing}
\end{figure}

Fig.(\ref{C_Tpairing}) compares the heat capacities of the SMSPS and the SMSPS with $\Delta=0$. Both heat capacities are computed using 11 orbits. In both isotopes the TVM peaks of SMSPS with pairing are shifted towards higher $T$, and the humps disappear. This is expected since increasing the energy gap due to pairing prevents neutrons in the Fermi level from melting at low $T$. The hump must have been shifted to higher $T$ and
disappeared under the TVM peak. The pairing effect increases the values of heat capacities
since a higher temperature is needed to overcome the correlation among valence nucleons.

\begin{figure}[ptbh]
\begin{center}
\includegraphics[width=0.90\textwidth]{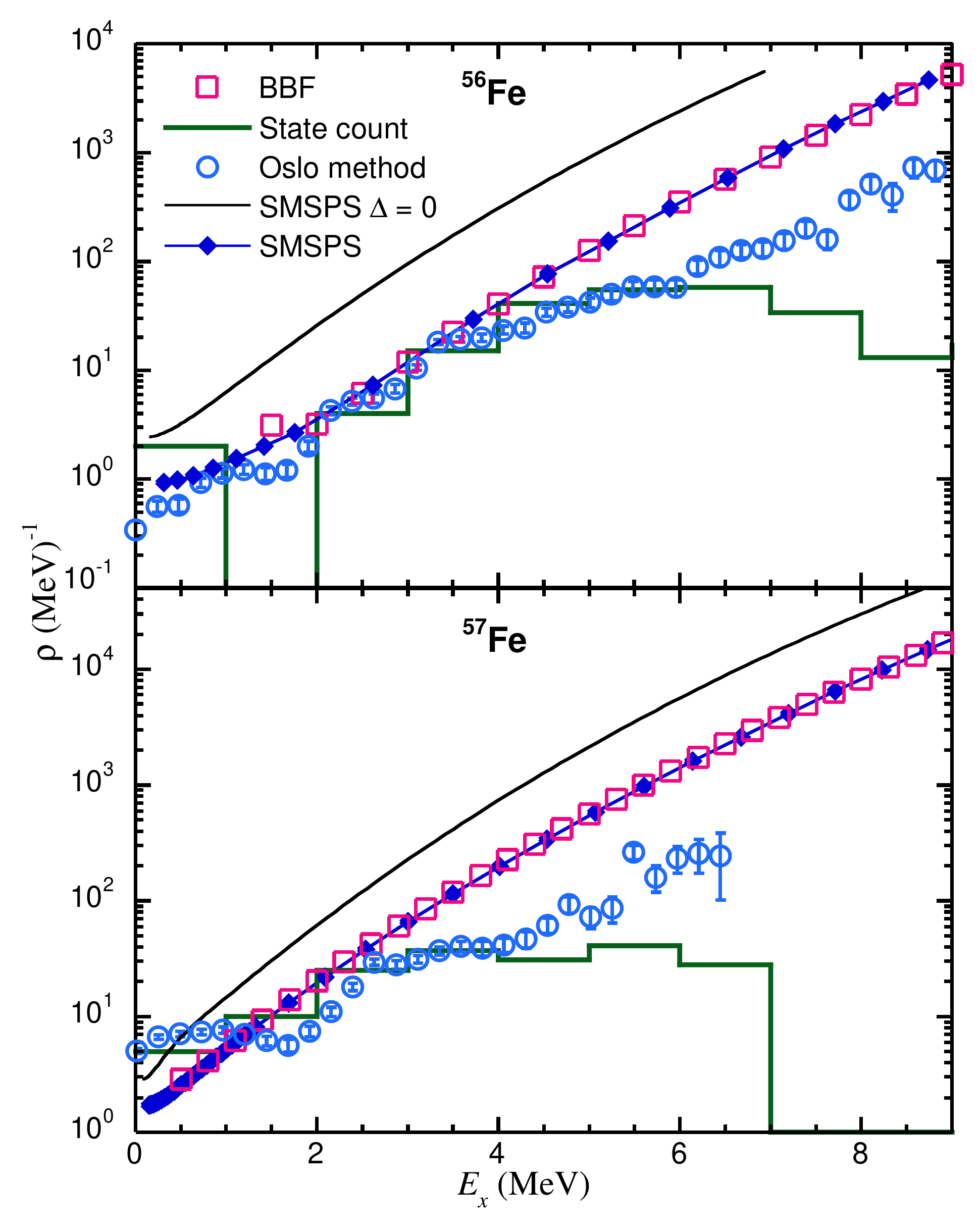}
\end{center}
\caption{(Color online) Same as Fig.(\ref{g_lowE}) but with paring included in the SMSPS level density of $^{56,57}$Fe. The SMSPS is computed at 4-orbit model space. The top and bottom figures have a common legend.}
\label{g_Epairing}
\end{figure}

\begin{figure}[ptbh]
\begin{center}
\includegraphics[width=0.90\textwidth]{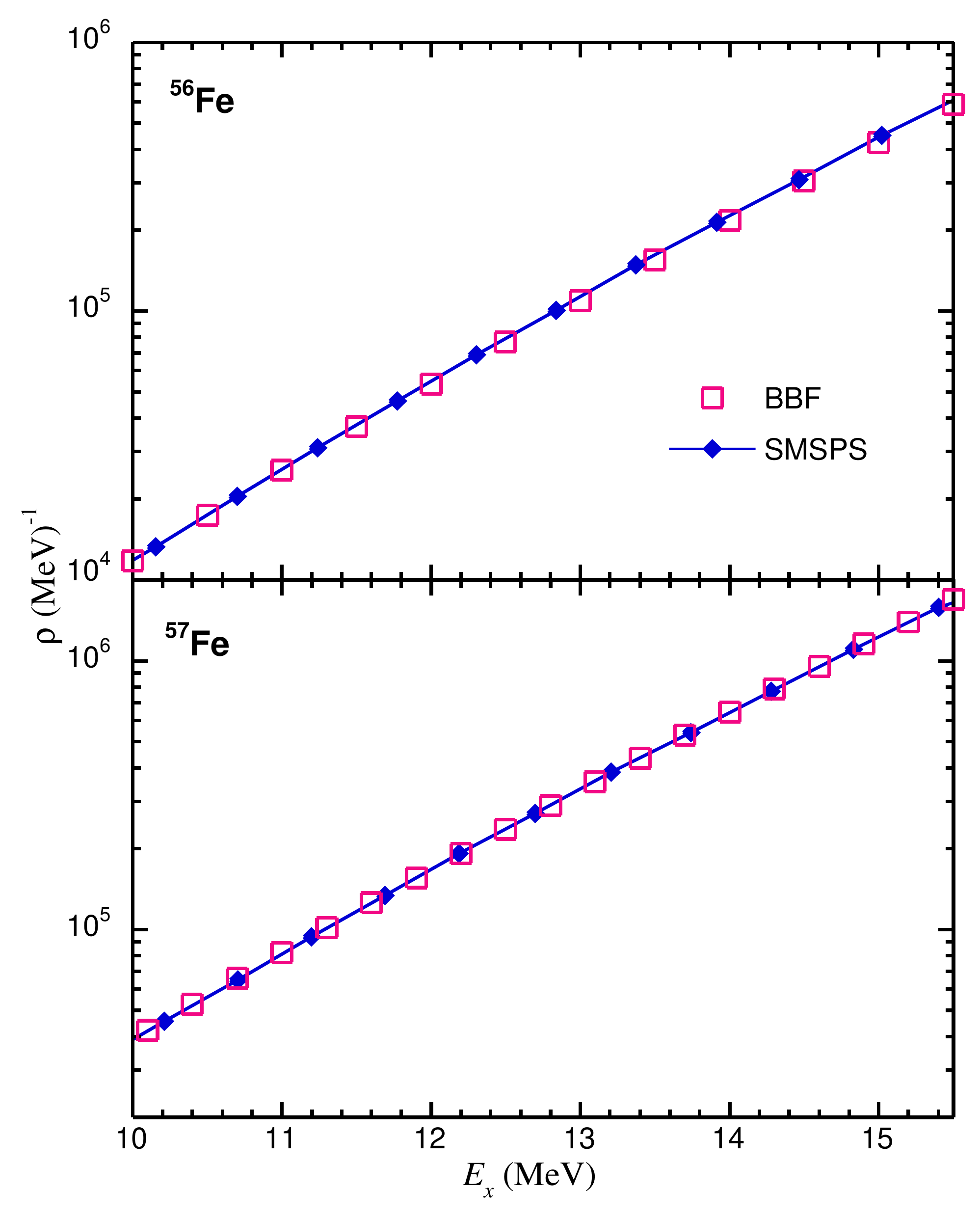}
\end{center}
\caption{(Color online) Same as Fig.(\ref{g_EHi}) with pairing included within the SMSPS high excitation level densities for $^{56,57}$Fe compared with the BBF. The SMSPS is computed with the 11-orbit model space.}
\label{g_Ehi_pairing}
\end{figure}

Fig.(\ref{g_Epairing}) shows the improvement that pairing brings to the SMSPS level
densities for both $^{56,57}$Fe. The agreement of the SMSPS with experimental data
and BBF is gratifying. This agreement with BBF for $^{56,57}$Fe extends to higher excitation energies, as Fig.(\ref{g_Ehi_pairing}) shows, beyond energies we used to evaluate $\Delta$ as a function of $T$.

\section{Conclusions and Outlook\label{secconclusion}}

We have shown that the SMSPS is a flexible tool to compute the thermal properties of isotopes in the Iron region. The SMSPS has the ability to produce results in both near zero and high temperature at large model spaces. Two primary variables needed to be adjusted in order to refine the SMSPS approach: (1) The volume of the continuum states which needs to be chosen such that the heat capacity exhibits a phase transition between quantum to classical gas at temperatures approximating the binding energy per nucleon, (2) The pairing parameter $\Delta$ which has to be calculated to achieve the correct backshift energy of the level density. The volume variable is crucial in the high-temperature region, whereas the pairing variable is important for even-even open-shell nuclei in the low temperature region. According to the SMSPS approach, $A = 56$ and $57$ Iron isotopes have the following properties
\begin{enumerate}
\item For $0<T< 0.9$ MeV these nuclei behave like a superfluid with significant pairing effects.
\item For $0.9<T<8.8$ MeV these nuclei exhibit properties reflective of a mixture of fluid (in the core) and quantum gas (in the valence and resonance states).
\item For $T>8.8$ MeV these nuclei exhibit the transition to a non-interacting classical gas.
\end{enumerate}

The main advantages of the SMSPS are:
\begin{enumerate}
\item Depends on the chosen SP states which means that the SMSPS accommodates physically realistic shell structures, including shell closure effects.
\item The theory is capable of treating all nucleons as active in a model space that extends to continuum.
\item The theory can admit pairing correlations in a phenomenological manner.
\item The SMSPS theory is computationally inexpensive.
\end{enumerate}
The disadvantages of the SMSPS are its sensitivity to the mean field SP states, and its
lack of a microscopic Hamiltonian. However, if the energy gap function due to the paring
correlation of such SPS is obtained and the volume parameter of the continuum state is
determined, the SMSPS is found to yield excellent values of level densities. In the future, we will investigate alternative models for the SPS such as those from mean field theory. The SMSPS model may be further improved by refining the configuration-restricted recursion technique with input from a fully microscopic NN interaction. However, this requires significant additional developments that we plan to report in the near future.

\begin{acknowledgments}
We thank S. Rombouts for providing us with his $^{56}$Fe results. Special
thanks are due to Y. Hida and D. Bailey for the help and support they
provided in constructing the multi-precision code used to evaluate the low
temperature observables. This work was supported in part by US\ Department
of Energy grant DE-FG-02-87ER40371. One of the authors B.A. Shehadeh acknowledges the support of Nuclear Physics server administrators at the Department of Physics and Astronomy at Iowa State University and the major support of the computation center at the Physics Department in Qassim University.
\end{acknowledgments}

\bibliographystyle{apsrev4-1}
\bibliography{basreferences}

\begin{thebibliography}{26}%
\makeatletter
\providecommand \@ifxundefined [1]{%
 \@ifx{#1\undefined}
}%
\providecommand \@ifnum [1]{%
 \ifnum #1\expandafter \@firstoftwo
 \else \expandafter \@secondoftwo
 \fi
}%
\providecommand \@ifx [1]{%
 \ifx #1\expandafter \@firstoftwo
 \else \expandafter \@secondoftwo
 \fi
}%
\providecommand \natexlab [1]{#1}%
\providecommand \enquote  [1]{``#1''}%
\providecommand \bibnamefont  [1]{#1}%
\providecommand \bibfnamefont [1]{#1}%
\providecommand \citenamefont [1]{#1}%
\providecommand \href@noop [0]{\@secondoftwo}%
\providecommand \href [0]{\begingroup \@sanitize@url \@href}%
\providecommand \@href[1]{\@@startlink{#1}\@@href}%
\providecommand \@@href[1]{\endgroup#1\@@endlink}%
\providecommand \@sanitize@url [0]{\catcode `\\12\catcode `\$12\catcode
  `\&12\catcode `\#12\catcode `\^12\catcode `\_12\catcode `\%12\relax}%
\providecommand \@@startlink[1]{}%
\providecommand \@@endlink[0]{}%
\providecommand \url  [0]{\begingroup\@sanitize@url \@url }%
\providecommand \@url [1]{\endgroup\@href {#1}{\urlprefix }}%
\providecommand \urlprefix  [0]{URL }%
\providecommand \Eprint [0]{\href }%
\providecommand \doibase [0]{http://dx.doi.org/}%
\providecommand \selectlanguage [0]{\@gobble}%
\providecommand \bibinfo  [0]{\@secondoftwo}%
\providecommand \bibfield  [0]{\@secondoftwo}%
\providecommand \translation [1]{[#1]}%
\providecommand \BibitemOpen [0]{}%
\providecommand \bibitemStop [0]{}%
\providecommand \bibitemNoStop [0]{.\EOS\space}%
\providecommand \EOS [0]{\spacefactor3000\relax}%
\providecommand \BibitemShut  [1]{\csname bibitem#1\endcsname}%
\let\auto@bib@innerbib\@empty
\bibitem [{\citenamefont {Alhassid}\ \emph {et~al.}(2015)\citenamefont
  {Alhassid}, \citenamefont {Bonett-Matiz}, \citenamefont {Liu},\ and\
  \citenamefont {Nakada}}]{ABLNphysrevc15}%
  \BibitemOpen
  \bibfield  {author} {\bibinfo {author} {\bibfnamefont {Y.}~\bibnamefont
  {Alhassid}}, \bibinfo {author} {\bibfnamefont {M.}~\bibnamefont
  {Bonett-Matiz}}, \bibinfo {author} {\bibfnamefont {S.}~\bibnamefont {Liu}}, \
  and\ \bibinfo {author} {\bibfnamefont {H.}~\bibnamefont {Nakada}},\ }\href
  {\doibase 10.1103/PhysRevC.92.024307} {\bibfield  {journal} {\bibinfo
  {journal} {Phys. Rev. C}\ }\textbf {\bibinfo {volume} {92}},\ \bibinfo
  {pages} {024307} (\bibinfo {year} {2015})}\BibitemShut {NoStop}%
\bibitem [{\citenamefont {Rombouts}\ \emph {et~al.}(1998)\citenamefont
  {Rombouts}, \citenamefont {Heyde},\ and\ \citenamefont
  {Jachowicz}}]{RHJphysrevc98}%
  \BibitemOpen
  \bibfield  {author} {\bibinfo {author} {\bibfnamefont {S.}~\bibnamefont
  {Rombouts}}, \bibinfo {author} {\bibfnamefont {K.}~\bibnamefont {Heyde}}, \
  and\ \bibinfo {author} {\bibfnamefont {N.}~\bibnamefont {Jachowicz}},\ }\href
  {\doibase 10.1103/PhysRevC.58.3295} {\bibfield  {journal} {\bibinfo
  {journal} {Phys. Rev. C}\ }\textbf {\bibinfo {volume} {58}},\ \bibinfo
  {pages} {3295} (\bibinfo {year} {1998})}\BibitemShut {NoStop}%
\bibitem [{\citenamefont {Nakada}\ and\ \citenamefont
  {Alhassid}(1998)}]{NAphyslett98}%
  \BibitemOpen
  \bibfield  {author} {\bibinfo {author} {\bibfnamefont {H.}~\bibnamefont
  {Nakada}}\ and\ \bibinfo {author} {\bibfnamefont {Y.}~\bibnamefont
  {Alhassid}},\ }\href {\doibase 10.1016/s0370-2693(98)00911-3} {\bibfield
  {journal} {\bibinfo  {journal} {Phys. Lett. B}\ }\textbf {\bibinfo {volume}
  {436}},\ \bibinfo {pages} {231} (\bibinfo {year} {1998})}\BibitemShut
  {NoStop}%
\bibitem [{\citenamefont {Liu}\ \emph {et~al.}(2000)\citenamefont {Liu},
  \citenamefont {Alhassid},\ and\ \citenamefont {Nakada}}]{LANaip00}%
  \BibitemOpen
  \bibfield  {author} {\bibinfo {author} {\bibfnamefont {S.}~\bibnamefont
  {Liu}}, \bibinfo {author} {\bibfnamefont {Y.}~\bibnamefont {Alhassid}}, \
  and\ \bibinfo {author} {\bibfnamefont {H.}~\bibnamefont {Nakada}},\
  }\href@noop {} {\bibfield  {journal} {\bibinfo  {journal} {Proc. Tenth Int.
  Symp. Capture gamma-ray Spec. S. Wender, ed. AIP conference proceedings}\ }
  (\bibinfo {year} {2000})}\BibitemShut {NoStop}%
\bibitem [{\citenamefont {Nakada}\ and\ \citenamefont
  {Alhassid}(1997)}]{NAphysrevlett97}%
  \BibitemOpen
  \bibfield  {author} {\bibinfo {author} {\bibfnamefont {H.}~\bibnamefont
  {Nakada}}\ and\ \bibinfo {author} {\bibfnamefont {Y.}~\bibnamefont
  {Alhassid}},\ }\href {\doibase 10.1103/PhysRevLett.79.2939} {\bibfield
  {journal} {\bibinfo  {journal} {Phys. Rev. Lett.}\ }\textbf {\bibinfo
  {volume} {79}},\ \bibinfo {pages} {2939} (\bibinfo {year}
  {1997})}\BibitemShut {NoStop}%
\bibitem [{\citenamefont {Firestone}\ and\ \citenamefont
  {Baglin}(1998)}]{toi98}%
  \BibitemOpen
  \bibfield  {author} {\bibinfo {author} {\bibfnamefont {R.~B.}\ \bibnamefont
  {Firestone}}\ and\ \bibinfo {author} {\bibfnamefont {C.}~\bibnamefont
  {Baglin}},\ }\href {https://books.google.com.sa/books?id=j1l7swEACAAJ} {\emph
  {\bibinfo {title} {Table of Isotopes, 2 Volume Set, 1998 Update}}},\ \bibinfo
  {series} {A Wiley-Interscience publication}\ No.\ \bibinfo {number} {v. 1-2}\
  (\bibinfo  {publisher} {Wiley},\ \bibinfo {year} {1998})\BibitemShut
  {NoStop}%
\bibitem [{\citenamefont {Schiller}\ \emph {et~al.}(2003)\citenamefont
  {Schiller}, \citenamefont {Algin}, \citenamefont {Bernstein}, \citenamefont
  {Garrett}, \citenamefont {Guttormsen}, \citenamefont {Hjorth-Jensen},
  \citenamefont {Johnson}, \citenamefont {Mitchell}, \citenamefont {Rekstad},
  \citenamefont {Siem}, \citenamefont {Voinov},\ and\ \citenamefont
  {Younes}}]{SETALphysRev03}%
  \BibitemOpen
  \bibfield  {author} {\bibinfo {author} {\bibfnamefont {A.}~\bibnamefont
  {Schiller}}, \bibinfo {author} {\bibfnamefont {E.}~\bibnamefont {Algin}},
  \bibinfo {author} {\bibfnamefont {L.}~\bibnamefont {Bernstein}}, \bibinfo
  {author} {\bibfnamefont {P.}~\bibnamefont {Garrett}}, \bibinfo {author}
  {\bibfnamefont {M.}~\bibnamefont {Guttormsen}}, \bibinfo {author}
  {\bibfnamefont {M.}~\bibnamefont {Hjorth-Jensen}}, \bibinfo {author}
  {\bibfnamefont {C.}~\bibnamefont {Johnson}}, \bibinfo {author} {\bibfnamefont
  {G.}~\bibnamefont {Mitchell}}, \bibinfo {author} {\bibfnamefont
  {J.}~\bibnamefont {Rekstad}}, \bibinfo {author} {\bibfnamefont
  {S.}~\bibnamefont {Siem}}, \bibinfo {author} {\bibfnamefont {A.}~\bibnamefont
  {Voinov}}, \ and\ \bibinfo {author} {\bibfnamefont {W.}~\bibnamefont
  {Younes}},\ }\href {\doibase 10.1103/PhysRevC.68.054326} {\bibfield
  {journal} {\bibinfo  {journal} {Phys. Rev. C}\ }\textbf {\bibinfo {volume}
  {68}},\ \bibinfo {pages} {054326} (\bibinfo {year} {2003})}\BibitemShut
  {NoStop}%
\bibitem [{\citenamefont {Schiller}\ \emph {et~al.}(2000)\citenamefont
  {Schiller}, \citenamefont {Bergholt}, \citenamefont {Guttormsen},
  \citenamefont {Melby}, \citenamefont {Rekstad},\ and\ \citenamefont
  {Siem}}]{SETALNulcInstMethA00}%
  \BibitemOpen
  \bibfield  {author} {\bibinfo {author} {\bibfnamefont {A.}~\bibnamefont
  {Schiller}}, \bibinfo {author} {\bibfnamefont {L.}~\bibnamefont {Bergholt}},
  \bibinfo {author} {\bibfnamefont {M.}~\bibnamefont {Guttormsen}}, \bibinfo
  {author} {\bibfnamefont {E.}~\bibnamefont {Melby}}, \bibinfo {author}
  {\bibfnamefont {J.}~\bibnamefont {Rekstad}}, \ and\ \bibinfo {author}
  {\bibfnamefont {S.}~\bibnamefont {Siem}},\ }\href {\doibase
  10.1016/s0168-9002(99)01187-0} {\bibfield  {journal} {\bibinfo  {journal}
  {Nucl. Inst. Meth. Phys. Res. A}\ }\textbf {\bibinfo {volume} {447}},\
  \bibinfo {pages} {498} (\bibinfo {year} {2000})}\BibitemShut {NoStop}%
\bibitem [{\citenamefont {Shlomo}\ and\ \citenamefont
  {J.B.Natowitz}(1990)}]{SNphyslettb90}%
  \BibitemOpen
  \bibfield  {author} {\bibinfo {author} {\bibfnamefont {S.}~\bibnamefont
  {Shlomo}}\ and\ \bibinfo {author} {\bibnamefont {J.B.Natowitz}},\ }\href
  {\doibase 10.1016/0370-2693(90)90859-5} {\bibfield  {journal} {\bibinfo
  {journal} {Phys. Lett. B}\ }\textbf {\bibinfo {volume} {252}},\ \bibinfo
  {pages} {187} (\bibinfo {year} {1990})}\BibitemShut {NoStop}%
\bibitem [{\citenamefont {Paar}\ and\ \citenamefont
  {Pezer}(1997)}]{PPphysrevc97}%
  \BibitemOpen
  \bibfield  {author} {\bibinfo {author} {\bibfnamefont {V.}~\bibnamefont
  {Paar}}\ and\ \bibinfo {author} {\bibfnamefont {R.}~\bibnamefont {Pezer}},\
  }\href {\doibase 10.1103/PhysRevC.55.R1637} {\bibfield  {journal} {\bibinfo
  {journal} {Phys. Rev. C}\ }\textbf {\bibinfo {volume} {55}},\ \bibinfo
  {pages} {R1637} (\bibinfo {year} {1997})}\BibitemShut {NoStop}%
\bibitem [{\citenamefont {Pratt}(2000)}]{Pphysrevlett00}%
  \BibitemOpen
  \bibfield  {author} {\bibinfo {author} {\bibfnamefont {S.}~\bibnamefont
  {Pratt}},\ }\href {\doibase 10.1103/PhysRevLett.84.4255} {\bibfield
  {journal} {\bibinfo  {journal} {Phys. Rev. Lett.}\ }\textbf {\bibinfo
  {volume} {84}},\ \bibinfo {pages} {4255} (\bibinfo {year}
  {2000})}\BibitemShut {NoStop}%
\bibitem [{\citenamefont {Chasman}(2003)}]{chasman03}%
  \BibitemOpen
  \bibfield  {author} {\bibinfo {author} {\bibfnamefont {R.}~\bibnamefont
  {Chasman}},\ }\href {\doibase https://doi.org/10.1016/S0370-2693(02)03226-4}
  {\bibfield  {journal} {\bibinfo  {journal} {Phys. Lett. B}\ }\textbf
  {\bibinfo {volume} {553}},\ \bibinfo {pages} {204 } (\bibinfo {year}
  {2003})}\BibitemShut {NoStop}%
\bibitem [{\citenamefont {Alhassid}\ \emph {et~al.}(2016)\citenamefont
  {Alhassid}, \citenamefont {Bertsch}, \citenamefont {Gilbreth},\ and\
  \citenamefont {Nakada}}]{PhysRevC.93.044320}%
  \BibitemOpen
  \bibfield  {author} {\bibinfo {author} {\bibfnamefont {Y.}~\bibnamefont
  {Alhassid}}, \bibinfo {author} {\bibfnamefont {G.}~\bibnamefont {Bertsch}},
  \bibinfo {author} {\bibfnamefont {C.}~\bibnamefont {Gilbreth}}, \ and\
  \bibinfo {author} {\bibfnamefont {H.}~\bibnamefont {Nakada}},\ }\href
  {\doibase 10.1103/PhysRevC.93.044320} {\bibfield  {journal} {\bibinfo
  {journal} {Phys. Rev. C}\ }\textbf {\bibinfo {volume} {93}},\ \bibinfo
  {pages} {044320} (\bibinfo {year} {2016})}\BibitemShut {NoStop}%
\bibitem [{\citenamefont {Martin}\ \emph {et~al.}(2003)\citenamefont {Martin},
  \citenamefont {Egido},\ and\ \citenamefont {Robledo}}]{MER_Physrev03}%
  \BibitemOpen
  \bibfield  {author} {\bibinfo {author} {\bibfnamefont {V.}~\bibnamefont
  {Martin}}, \bibinfo {author} {\bibfnamefont {J.}~\bibnamefont {Egido}}, \
  and\ \bibinfo {author} {\bibfnamefont {L.}~\bibnamefont {Robledo}},\ }\href
  {\doibase 10.1103/PhysRevC.68.034327} {\bibfield  {journal} {\bibinfo
  {journal} {Phys. Rev. C}\ }\textbf {\bibinfo {volume} {68}},\ \bibinfo
  {pages} {034327} (\bibinfo {year} {2003})}\BibitemShut {NoStop}%
\bibitem [{\citenamefont {Khan}\ \emph {et~al.}(2007)\citenamefont {Khan},
  \citenamefont {Giai},\ and\ \citenamefont {Sandulescu}}]{KVS_NuclPhysA07}%
  \BibitemOpen
  \bibfield  {author} {\bibinfo {author} {\bibfnamefont {E.}~\bibnamefont
  {Khan}}, \bibinfo {author} {\bibfnamefont {N.~V.}\ \bibnamefont {Giai}}, \
  and\ \bibinfo {author} {\bibfnamefont {N.}~\bibnamefont {Sandulescu}},\
  }\href {\doibase https://doi.org/10.1016/j.nuclphysa.2007.03.005} {\bibfield
  {journal} {\bibinfo  {journal} {Nuclear Physics A}\ }\textbf {\bibinfo
  {volume} {789}},\ \bibinfo {pages} {94 } (\bibinfo {year}
  {2007})}\BibitemShut {NoStop}%
\bibitem [{\citenamefont {Gambacurta}\ \emph {et~al.}(2013)\citenamefont
  {Gambacurta}, \citenamefont {Lacroix},\ and\ \citenamefont
  {Sandulescu}}]{PhysRevC.88.034324}%
  \BibitemOpen
  \bibfield  {author} {\bibinfo {author} {\bibfnamefont {D.}~\bibnamefont
  {Gambacurta}}, \bibinfo {author} {\bibfnamefont {D.}~\bibnamefont {Lacroix}},
  \ and\ \bibinfo {author} {\bibfnamefont {N.}~\bibnamefont {Sandulescu}},\
  }\href {\doibase 10.1103/PhysRevC.88.034324} {\bibfield  {journal} {\bibinfo
  {journal} {Phys. Rev. C}\ }\textbf {\bibinfo {volume} {88}},\ \bibinfo
  {pages} {034324} (\bibinfo {year} {2013})}\BibitemShut {NoStop}%
\bibitem [{\citenamefont {Gambacurta}\ \emph {et~al.}(2014)\citenamefont
  {Gambacurta}, \citenamefont {Lacroix},\ and\ \citenamefont
  {Sandulescu}}]{GLS_PSH_2014}%
  \BibitemOpen
  \bibfield  {author} {\bibinfo {author} {\bibfnamefont {D.}~\bibnamefont
  {Gambacurta}}, \bibinfo {author} {\bibfnamefont {D.}~\bibnamefont {Lacroix}},
  \ and\ \bibinfo {author} {\bibfnamefont {N.}~\bibnamefont {Sandulescu}},\
  }\href {http://stacks.iop.org/1742-6596/533/i=1/a=012012} {\bibfield
  {journal} {\bibinfo  {journal} {Journal of Physics: Conference Series}\
  }\textbf {\bibinfo {volume} {533}},\ \bibinfo {pages} {012012} (\bibinfo
  {year} {2014})}\BibitemShut {NoStop}%
\bibitem [{\citenamefont {Borrmann}\ and\ \citenamefont
  {Franke}(1993)}]{BFjcp93}%
  \BibitemOpen
  \bibfield  {author} {\bibinfo {author} {\bibfnamefont {P.}~\bibnamefont
  {Borrmann}}\ and\ \bibinfo {author} {\bibfnamefont {G.}~\bibnamefont
  {Franke}},\ }\href {\doibase 10.1063/1.464180} {\bibfield  {journal}
  {\bibinfo  {journal} {Jour. Chem. Phys.}\ }\textbf {\bibinfo {volume} {98}},\
  \bibinfo {pages} {2484} (\bibinfo {year} {1993})}\BibitemShut {NoStop}%
\bibitem [{\citenamefont {Fowler}\ \emph {et~al.}(1978)\citenamefont {Fowler},
  \citenamefont {Engelbrecht},\ and\ \citenamefont
  {Woosley}}]{FEW_AstroJour78}%
  \BibitemOpen
  \bibfield  {author} {\bibinfo {author} {\bibfnamefont {W.}~\bibnamefont
  {Fowler}}, \bibinfo {author} {\bibfnamefont {C.}~\bibnamefont {Engelbrecht}},
  \ and\ \bibinfo {author} {\bibfnamefont {S.}~\bibnamefont {Woosley}},\
  }\href@noop {} {\bibfield  {journal} {\bibinfo  {journal} {Astro. Jour.}\
  }\textbf {\bibinfo {volume} {226}},\ \bibinfo {pages} {984} (\bibinfo {year}
  {1978})}\BibitemShut {NoStop}%
\bibitem [{\citenamefont {Shehadeh}(2003)}]{phd}%
  \BibitemOpen
  \bibfield  {author} {\bibinfo {author} {\bibfnamefont {B.}~\bibnamefont
  {Shehadeh}},\ }\emph {\bibinfo {title} {Statistical and thermal properties of
  mesoscopic systems: Application to the many-nucleon system.}},\ \href@noop {}
  {Ph.D. thesis},\ \bibinfo  {school} {Iowa State University} (\bibinfo {year}
  {2003})\BibitemShut {NoStop}%
\bibitem [{\citenamefont {Hida}\ \emph {et~al.}(2001)\citenamefont {Hida},
  \citenamefont {Li},\ and\ \citenamefont {Bailey}}]{hida1}%
  \BibitemOpen
  \bibfield  {author} {\bibinfo {author} {\bibfnamefont {Y.}~\bibnamefont
  {Hida}}, \bibinfo {author} {\bibfnamefont {X.}~\bibnamefont {Li}}, \ and\
  \bibinfo {author} {\bibfnamefont {D.}~\bibnamefont {Bailey}},\ }\href
  {https://pubarchive.lbl.gov/islandora/object/ir%3A117922} {\bibfield
  {journal} {\bibinfo  {journal} {Proceedings of the 15th IEEE Symposium on
  Computer Arithmetic (ARITH15)}\ } (\bibinfo {year} {2001})}\BibitemShut
  {NoStop}%
\bibitem [{\citenamefont {Hida}\ \emph {et~al.}(2000)\citenamefont {Hida},
  \citenamefont {Li},\ and\ \citenamefont {Bailey}}]{hida2}%
  \BibitemOpen
  \bibfield  {author} {\bibinfo {author} {\bibfnamefont {Y.}~\bibnamefont
  {Hida}}, \bibinfo {author} {\bibfnamefont {X.~S.}\ \bibnamefont {Li}}, \ and\
  \bibinfo {author} {\bibfnamefont {D.~H.}\ \bibnamefont {Bailey}},\
  }\href@noop {} {\emph {\bibinfo {title} {Quad-Double Arithmetic: Algorithms,
  Implementation, and Application}}},\ \bibinfo {type} {Tech. Rep.}\ (\bibinfo
  {institution} {Lawrence Berkeley National Laboratory, Berkeley, CA 94720},\
  \bibinfo {year} {2000})\BibitemShut {NoStop}%
\bibitem [{\citenamefont {Bailey}\ \emph {et~al.}(2018)\citenamefont {Bailey},
  \citenamefont {Hida}, \citenamefont {Li}, \citenamefont {Thompson},
  \citenamefont {Jeyabalan},\ and\ \citenamefont {Kaiser}}]{arprec}%
  \BibitemOpen
  \bibfield  {author} {\bibinfo {author} {\bibfnamefont {D.~H.}\ \bibnamefont
  {Bailey}}, \bibinfo {author} {\bibfnamefont {Y.}~\bibnamefont {Hida}},
  \bibinfo {author} {\bibfnamefont {X.~S.}\ \bibnamefont {Li}}, \bibinfo
  {author} {\bibfnamefont {B.}~\bibnamefont {Thompson}}, \bibinfo {author}
  {\bibfnamefont {K.}~\bibnamefont {Jeyabalan}}, \ and\ \bibinfo {author}
  {\bibfnamefont {A.}~\bibnamefont {Kaiser}},\ }\href
  {http://crd-legacy.lbl.gov/~dhbailey/mpdist/} {\enquote {\bibinfo {title}
  {High-precision software directory},}\ } (\bibinfo {year} {2018})\BibitemShut
  {NoStop}%
\bibitem [{\citenamefont {Gilbert}\ and\ \citenamefont
  {Cameron}(1965)}]{GCcanjphys65}%
  \BibitemOpen
  \bibfield  {author} {\bibinfo {author} {\bibfnamefont {A.}~\bibnamefont
  {Gilbert}}\ and\ \bibinfo {author} {\bibfnamefont {A.}~\bibnamefont
  {Cameron}},\ }\href {\doibase 10.1139/p65-139} {\bibfield  {journal}
  {\bibinfo  {journal} {Can. Jour. Phys.}\ }\textbf {\bibinfo {volume} {43}},\
  \bibinfo {pages} {1446} (\bibinfo {year} {1965})}\BibitemShut {NoStop}%
\bibitem [{\citenamefont {Woosley}\ \emph {et~al.}(1978)\citenamefont
  {Woosley}, \citenamefont {Fowler}, \citenamefont {Holmes},\ and\
  \citenamefont {Zimmerman}}]{WFHZatomnucldatatab78}%
  \BibitemOpen
  \bibfield  {author} {\bibinfo {author} {\bibfnamefont {S.}~\bibnamefont
  {Woosley}}, \bibinfo {author} {\bibfnamefont {W.}~\bibnamefont {Fowler}},
  \bibinfo {author} {\bibfnamefont {J.}~\bibnamefont {Holmes}}, \ and\ \bibinfo
  {author} {\bibfnamefont {B.}~\bibnamefont {Zimmerman}},\ }\href {\doibase
  https://doi.org/10.1016/0092-640X(78)90018-9} {\bibfield  {journal} {\bibinfo
   {journal} {Atomic Data and Nuclear Data Tables}\ }\textbf {\bibinfo {volume}
  {22}},\ \bibinfo {pages} {371 } (\bibinfo {year} {1978})}\BibitemShut
  {NoStop}%
\bibitem [{\citenamefont {Natowitz}\ \emph {et~al.}(2002)\citenamefont
  {Natowitz}, \citenamefont {Hagel}, \citenamefont {Ma}, \citenamefont
  {Murray}, \citenamefont {Qin}, \citenamefont {Shlomo}, \citenamefont {Wada},\
  and\ \citenamefont {Wang}}]{NHMMQSWWphysrevc02}%
  \BibitemOpen
  \bibfield  {author} {\bibinfo {author} {\bibfnamefont {J.~B.}\ \bibnamefont
  {Natowitz}}, \bibinfo {author} {\bibfnamefont {K.}~\bibnamefont {Hagel}},
  \bibinfo {author} {\bibfnamefont {Y.}~\bibnamefont {Ma}}, \bibinfo {author}
  {\bibfnamefont {M.}~\bibnamefont {Murray}}, \bibinfo {author} {\bibfnamefont
  {L.}~\bibnamefont {Qin}}, \bibinfo {author} {\bibfnamefont {S.}~\bibnamefont
  {Shlomo}}, \bibinfo {author} {\bibfnamefont {R.}~\bibnamefont {Wada}}, \ and\
  \bibinfo {author} {\bibfnamefont {J.}~\bibnamefont {Wang}},\ }\href {\doibase
  10.1103/PhysRevC.66.031601} {\bibfield  {journal} {\bibinfo  {journal} {Phys.
  Rev. C}\ }\textbf {\bibinfo {volume} {66}},\ \bibinfo {pages} {031601}
  (\bibinfo {year} {2002})}\BibitemShut {NoStop}%
\end{thebibliography}%

\end{document}